\documentclass[aps,prb,twocolumn,superscriptaddress]{revtex4-1}

\usepackage{graphicx}
\usepackage{amsmath}
\usepackage{mathtools}
\usepackage{color}

\usepackage{listings}

\lstdefinestyle{mystyle}{
    commentstyle=\color{black},
    captionpos=b,                    
    numbers=left,                    
    numbersep=5pt,                  
    showspaces=false,                
    showstringspaces=false,
}

\begin{document}

% Use the \preprint command to place your local institutional report number 
% on the title page in preprint mode.
% Multiple \preprint commands are allowed.
%\preprint{}

%\title{Practical $GW$ scheme for electronic structure of 3$d$-transition-metal monoxide anions: ScO$^{-}$, TiO$^{-}$, CuO$^{-}$, and ZnO$^{-}$} %Title of paper
%\title{Practical self-consistent $GW$ scheme for electronic structure of open-shell 3$d$-transition-metal monoxide anions} %Title of paper
%\title{GPU accelerating MOLGW with OpenACC} %Title of paper
\title{GPU acceleration of many-body perturbation theory methods in MOLGW with OpenACC} %Title of paper

% repeat the \author .. \affiliation  etc. as needed
% \email, \thanks, \homepage, \altaffiliation all apply to the current author.
% Explanatory text should go in the []'s, 
% actual e-mail address or url should go in the {}'s for \email and \homepage.
% Please use the appropriate macro for the type of information

% \affiliation command applies to all authors since the last \affiliation command. 
% The \affiliation command should follow the other information.

%\author{Young-Moo Byun and Serdar \"{O}\u{g}\"{u}t}
\author{Young-Moo Byun}
%\email[]{Your e-mail address}
%\homepage[]{Your web page}
%\thanks{}
%\altaffiliation{}
\affiliation{Department of Physics, Sungkyunkwan University, Suwon-si, Gyeonggi-do 16419, Korea}
%\affiliation{Department of Physics, University of Illinois at Chicago, Chicago, IL 60607, USA}

\author{Jejoong Yoo}
\email[]{jejoong@skku.edu}
\affiliation{Department of Physics, Sungkyunkwan University, Suwon-si, Gyeonggi-do 16419, Korea}

%\author{Serdar \"{O}\u{g}\"{u}t}
%\email[]{ogut@uic.edu}
%\affiliation{Department of Physics, University of Illinois at Chicago, Chicago, IL 60607, USA}

% Collaboration name, if desired (requires use of superscriptaddress option in \documentclass). 
% \noaffiliation is required (may also be used with the \author command).
%\collaboration{}
%\noaffiliation

\date{\today}

\begin{abstract}
% insert abstract here
Quasiparticle self-consistent many-body perturbation theory (MBPT) methods that update both eigenvalues and eigenvectors can calculate the excited-state properties of molecular systems without depending on the choice of starting points. However, those methods are computationally intensive even on modern multi-core central processing units (CPUs) and thus typically limited to small systems. Many-core accelerators such as graphics processing units (GPUs) may be able to boost the performance of those methods without losing accuracy, making starting-point-independent MBPT methods applicable to large systems. Here, we GPU accelerate MOLGW, a Gaussian-based MBPT code for molecules, with open accelerators (OpenACC) and achieve speedups of up to \textcolor{black}{9.7x} over 32 open multi-processing (OpenMP) CPU threads.
\end{abstract}

%10x
%16 open multi-processing (OpenMP) CPU threads
%on a personal computer
%13.6x

\pacs{}% insert suggested PACS numbers in braces on next line

\maketitle %\maketitle must follow title, authors, abstract and \pacs

% Body of paper goes here. Use proper sectioning commands. 
% References should be done using the \cite, \ref, and \label commands
%\section{}
%\label{}
%\subsection{}
%\subsubsection{}

\section{Introduction} \label{sec:Introduction}

%are a promising candidate

A theoretical method capable of accurately and efficiently describing the excited-state properties of molecular systems is important for the rational design of molecules with desired properties. Many-body perturbation theory (MBPT) methods based on the one-body Green's function $G$ \textcolor{black}{might be a good choice}, because the $GW$ method with $W$ being the dynamically screened Coulomb interaction has been successfully used to calculate electronic excitations in solids for a few decades.~\cite{Onida02} For example, the non-self-consistent (one-shot) $GW$ method ($G_{0}W_{0}$) starting from the Perdew--Burke--Ernzerhof (PBE) generalized gradient approximation to density-functional theory ($G_{0}W_{0}$@PBE using the \{MBPT method\}@\{starting point\} notation) can accurately calculate the bandgap of semiconductors at a reasonably low computational cost.~\cite{Shishkin07a,Govoni15}

%qs$GW$ calculations for large systems
%\textcolor{black}{of}

However, $G_{0}W_{0}$@PBE fails for molecules, significantly underestimating the ionization energy (IE) of small $sp$ and 3$d$ molecules by $\sim$0.5 and $\sim$1.0~eV, respectively.~\cite{vanSetten15,Byun19} A simple workaround to make the $G_{0}W_{0}$ method work for finite systems is to change its starting point from the PBE exchange-correlation functional to the Hartree--Fock (HF) exchange or hybrid functionals, which admix a fraction of non-local HF (exact) exchange with semilocal exchange.~\cite{Bruneval13} However, this workaround makes the $G_{0}W_{0}$ method \emph{empirical}; $G_{0}W_{0}$ is system dependent and its results depend strongly on the choice of starting points.~\cite{Byun19,Byun21} The quasiparticle self-consistent $GW$ (qs$GW$) method can address the system and starting-point dependency issues of the $G_{0}W_{0}$ method,~\cite{Faleev04,vanSchilfgaarde06,Kotani07} giving good results for both the bandgap of solids and the IE of atoms and molecules.~\cite{Bruneval14,Koval14,Kaplan16,Caruso16,Byun21,vanSetten18} However, the qs$GW$ method is computationally expensive even for small systems, making qs$GW$ calculations \textcolor{black}{of} large systems not feasible even on high-performance computing (HPC) supercomputers powered by modern multi-core central processing units (CPUs).

Graphics processing units (GPUs) may solve the efficiency problem of the qs$GW$ method, enabling large-scale qs$GW$ calculations without compromising accuracy.~\cite{Ben20,Barca21,Yu22} GPUs were accelerators (special-purpose hardware) designed for computer graphics at first, but later general-purpose computing on graphics processing units (GPGPU) made it possible for GPUs to perform non-specialized computations that have typically been conducted by CPUs. On the hardware side, GPUs consisting of many lightweight cores can accelerate numerically intensive and massively parallel computations in a single instruction multiple threads (SIMT) fashion. Some modern GPUs supporting high-performance double-precision floating-point (FP64) computations are widely used for HPC scientific applications that require high precision and accuracy for reliability and stability, such as \textit{ab initio} quantum chemistry applications.~\cite{Walker16a,Walker16b,Gotz16,DePrince11,Isborn11,Gordon20a,Gordon20b,Calvin21}

On the software side, there are two popular and mature GPU programming models: compute unified device architecture (CUDA) and open accelerators (OpenACC). While CUDA is a low-level model and is an extension to C/C++/Fortran,~\cite{CUDA} like open multi-processing (OpenMP), OpenACC is a high-level model based on compiler directives.~\cite{OpenACC} CUDA enables to harness the parallel computing power of GPUs, but CUDA programming \textcolor{black}{could be complex}, because it requires an understanding of the GPU hardware architecture and a significant modification of the original CPU source code. Also, the CUDA code runs only on NVIDIA GPUs. OpenACC is an alternative to CUDA to simplify parallel programming and make the parallel code portable across various kinds of platforms such as operating systems, compilers, CPUs, and GPUs.

%CUDA programming is difficult

\textcolor{black}{As the creator of the Python programming language said ``Maintainable code is more important than clever code,''~\cite{Dropbox} maintainability is as important as performance in software development, including the development of HPC scientific applications.} OpenACC achieves a balance between productivity and performance by enabling scientists and researchers to accelerate existing CPU codes on GPUs quickly with minimal programming effort, and thus is becoming increasingly popular.~\cite{Eriksen17,Bykov17,VergaraLarrea20,Smith22,Dang22} For example, the GPU port of VASP has lately transitioned from CUDA to OpenACC, because ``OpenACC dramatically decreases GPU development and maintenance efforts."~\cite{Maintz18a,Maintz18b}

%OpenACC is becoming increasingly popular, because it enables scientists and researchers to accelerate existing CPU codes on GPUs quickly with minimal programming effort and thereby achieves a balance between productivity and performance.~\cite{Eriksen17,Bykov17,VergaraLarrea20,Smith22,Dang22} For example, the GPU port of VASP has lately transitioned from CUDA to OpenACC, because ``OpenACC dramatically decreases GPU development and maintenance efforts."~\cite{Maintz18a,Maintz18b}

%First, as the creator of the Python programming language said ``Maintainable code is more important than clever code,''~\cite{Dropbox} maintainability is as important as performance in software development including the development of HPC scientific applications. OpenMP and OpenACC implementations in MOLGW ``added parallelism into existing source code without significantly modifying it,''~\cite{GuideOpenMP} enabling us to maintain a single version of source code for both CPUs and GPUs. The clean and maintainable MOLGW source code makes it possible for domain scientists, such as electronic structure method developers, with no or little parallel programming background to do ``more science and less programming.''~\cite{OpenMP}

In recent years, MBPT methods for molecules with local orbitals have been implemented in a variety of electronic structure codes, including FIESTA,~\cite{Blase11} FHI-AIMS,~\cite{Ren12} TURBOMOLE,~\cite{vanSetten13} CP2K,~\cite{Wilhelm16} and MOLGW.~\cite{Bruneval16} Among those codes, MOLGW is a double-precision Fortran/C++/Python CPU code for MBPT excited-state calculations of molecular systems based on Gaussian basis sets. Although one of us has recently parallelized MOLGW with OpenMP, starting-point-independent MBPT methods, such as qs$GW$, in OpenMP-parallelized MOLGW still are too time-consuming to be applied to large systems.~\cite{Byun19,Byun21}

In this work, we port MOLGW to the GPU using OpenACC to extend the range of its applicability while keeping the original CPU source code intact. We benchmark the performance of GPU-enabled MOLGW and find that \textcolor{black}{the OpenACC version of MOLGW on desktop GPUs can outperform the OpenMP version on workstation and supercomputer CPUs.}

%personal computer (PC)
%even on desktop GPUs
%using a workstation and a supercomputer

The rest of this paper is structured as follows: First, we give a mathematical introduction to MBPT. Second, we overview our OpenACC implementation of MBPT methods in MOLGW. Third, we analyze and discuss our benchmark results for the performance of our GPU-accelerated MOLGW. Last, we summarize and conclude.

\begin{figure*}
\begin{tabular}{c}
{\includegraphics[trim=0mm 0mm 0mm 0mm, clip, width=0.96\textwidth]{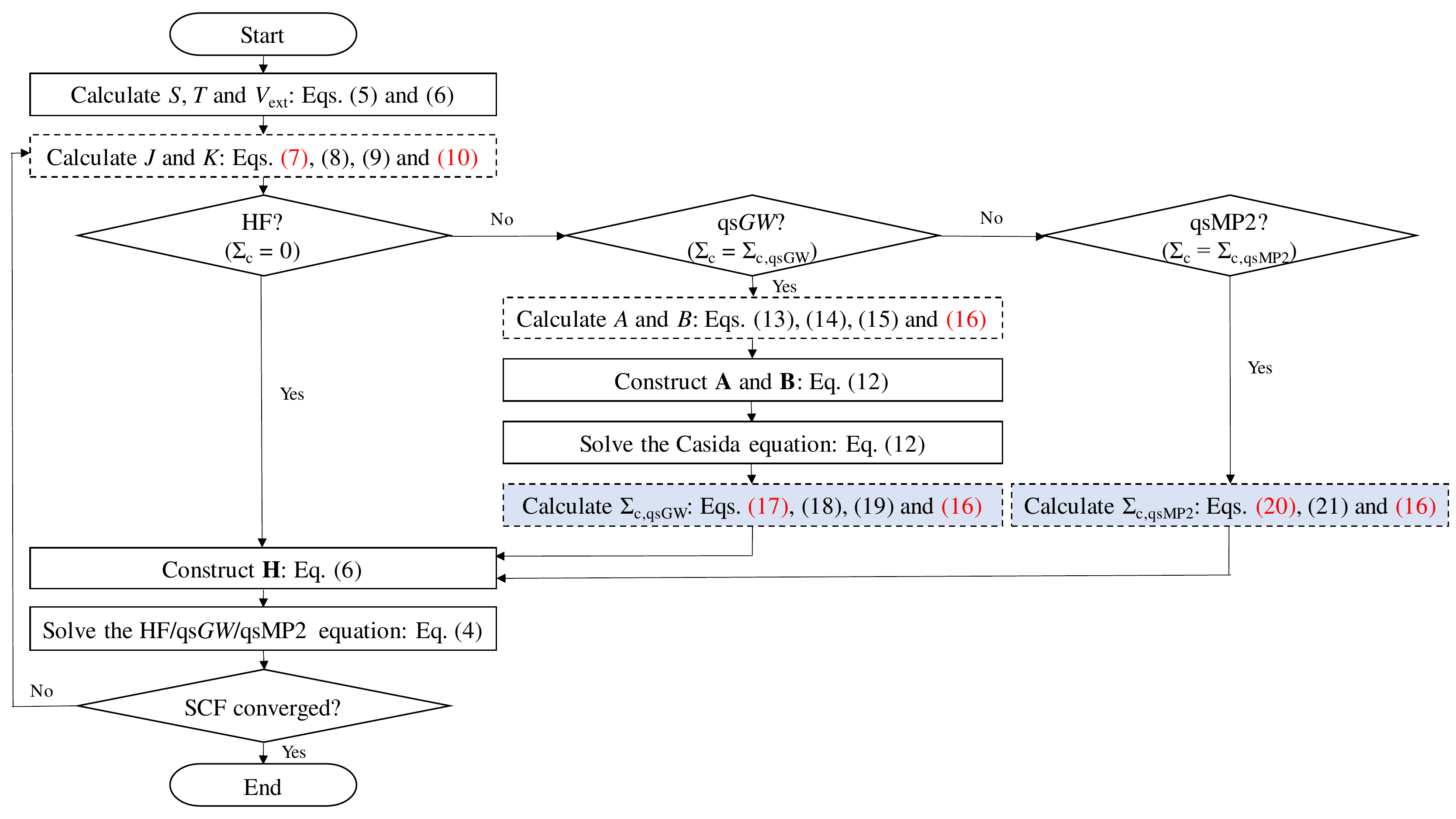}}
\end{tabular}
\caption{\textcolor{black}{(Color online) Flowchart of qs$GW$ and qsMP2 implementations in MOLGW. Boxes with a dashed line represent computational bottlenecks in this work. Light blue boxes represent major computational bottlenecks in this work. OpenMP- and OpenACC-parallelized equations are highlighted in red.}}
\label{fig:flowchart}
\end{figure*}

%qs$GW$ and qsMP2 methods implemented in MOLGW
%w/o

\begin{table*}
  \caption{\textcolor{black}{OpenACC directives and clauses used in this work. $N_{\text{kernel}}$ represents the number of kernels in a computational bottleneck. AO-MO represents the atomic orbital-to-molecular orbital integral transformation.}}
  \label{tab:OpenACC.directives.clauses}
  \begin{tabular*}{1.00\textwidth}{ @{\extracolsep{\fill}} l l r c l }
    \hline \hline
Bottleneck & Equations & Function/subroutine and file\footnotemark[1] & $N_{\text{kernel}}$ & OpenACC directives and clauses\footnotemark[2] \\
\hline
N/A\footnotemark[3]                         & Eq.~(\ref{eq:AO.4.ERI})  & See footnote\footnotemark[4]     & N/A\footnotemark[3] & declare/create/update/device \& \\
                                                         &                                       & m\_eri.f90                                  &       & routine/seq \\
\hline
Hartree $J$                                        & Eq.~(\ref{eq:J})             & setup\_hartree()                         & 1    & data/copyin/copy \& \\
                                                         &                                       & m\_hamiltonian.f90                     &       & parallel/loop/collapse/seq \\
\hline
Exchange $K$                                    & Eq.~(\ref{eq:K})             & setup\_exchange()                      & 1    & data/copyin/copy \& \\
                                                         &                                       & m\_hamiltonian.f90                     &       & parallel/loop/collapse/seq \\
\hline
AO-MO\footnotemark[5]                                               & Eq.~(\ref{eq:AO.to.MO}) & calculate\_eri\_4center\_eigen() & 4    & data/copyin/copyout \& \\
                                                         &                                       & m\_eri\_ao\_mo.f90                    &       & parallel/loop/collapse/seq \\
\hline
$\Sigma^{\text{qsGW}}_{\text{c}}$ excl. AO-MO\footnotemark[5]  & Eqs.~(\ref{eq:Sigma.c.GW}), (\ref{eq:w.residue}) \& (\ref{eq:Sigma.c.qsGW.real}) & chi\_to\_vchiv()           & 1 & data/enter data/exit data/copyin/copyout\footnotemark[6] \& \\
                                                                                       &                                                                                                                    & linear\_response.f90    &    & parallel/loop/collapse \\
\hline
$\Sigma^{\text{qsMP2}}_{\text{c}}$ excl. AO-MO\footnotemark[5] & Eqs.~(\ref{eq:Sigma.c.MP2}) \& (\ref{eq:Sigma.c.qsMP2.real})                                & pt2\_selfenergy\_qs()  & 1 & data/copyin/copy \& \\
                                                                                       &                                                                                                                    & pt2\_selfenergy.f90     &    & parallel/loop/collapse/seq \\
    \hline \hline
  \end{tabular*}
  \footnotetext[1]{Functions and subroutines are in the first line, and files are in the second line.}
  \footnotetext[2]{For data movement (first line) and computation (second line)}
  \footnotetext[3]{Not available because what OpenACC does in this Fortran module is not parallelizing computational bottlenecks, but transferring global data from CPU memory to GPU memory.}
  \footnotetext[4]{Functions: negligible\_basispair(), index\_eri(), index\_pair(), and eri(). Subroutines: prepare\_eri(), setup\_shell\_index(), setup\_basispair(), and identify\_negligible\_shellpair().}
  \footnotetext[5]{These three are major computational bottlenecks in this work.}
  \footnotetext[6]{CUDA unified (managed) memory can optimize data transfer in this kernel.}
\end{table*}

%quasiparticle (QP)
%qs$GW$ quasiparticle (QP) energies for the highest occupied molecular orbital (HOMO) of the closed-shell singlet CuO$^{-}$ anion during the self-consistent field (SCF) iteration, obtained from OpenMP and OpenACC calculations. The aug-cc-pVQZ basis set, $\eta$ = 0.05~Ha in Eq.~(\ref{eq:Sigma.c.GW}), and the calculated bond length of 1.697~{\AA} are used.
%[$\Sigma^{\sigma}_{\text{c}} = 0$ in Eq.~(\ref{eq:H})]
%[$\Sigma^{\sigma}_{\text{c}} = \Sigma^{\text{qsGW},\sigma}_{\text{c}}$ in Eq.~(\ref{eq:H})]

%pt2_selfenergy_qs(nstate,basis,occupation,energy,c_matrix,s_matrix,selfenergy,emp2)
%copyin(occupation,energy,eri_eigenstate_i)
%if( .NOT. has_auxil_basis ) then
%endif
%!LOOP of the first Green's function
%! external loop ( bra )
%! external loop ( ket )
%calculate_eri_4center_eigen(basis%nbf,nstate,c_matrix,istate,pqispin,eri_eigenstate_i)
%PRIVATE(pstate, qstate, jkspin, jstate,fj,ej, kstate,fk,ek,fact_occ1,fact_occ2,coul_ipkj,coul_iqjk,coul_ijkq,ep,eq,fact_real,fact_energy)
%! pqispin
%copy(emp2_ring,emp2_sox,selfenergy_ring,selfenergy_sox)
%REDUCTION(+:emp2_ring,emp2_sox)
%$\Sigma^{\text{qsMP2},\sigma}_{\text{c}}$
%consisting of 4 kernels
%OpenMP- and OpenACC-parallelized MOLGW

\section{Theoretical background}

In this section, we give a minimal \textcolor{black}{and simplified} introduction to MBPT related to OpenMP and OpenACC implementations in MOLGW. More details about MBPT can be found elsewhere.~\cite{Ren12,Golze19,Bruneval14,Bruneval21,Byun21} We use a simple and consistent notation: (i) we follow the notations in MOLGW implementation and application papers,~\cite{Bruneval16,Byun19,Byun21} (ii) we use Hartree atomic units, and (iii) we omit the complex conjugate notation, because we consider real wavefunctions.

\subsection{Many-body perturbation theory (MBPT)}

MBPT can be used to calculate electron addition and removal energies. In MBPT, the central quantity is the one-body Green's function:
\begin{equation}
G^{\sigma}(\mathbf{r},\mathbf{r'},\omega) = \sum_{i} \frac{\varphi^{\sigma}_{i}(\mathbf{r})\varphi^{\sigma}_{i}(\mathbf{r'})}{\omega - \epsilon^{\sigma}_{i} - i\eta} + \sum_{a} \frac{\varphi^{\sigma}_{a}(\mathbf{r})\varphi^{\sigma}_{a}(\mathbf{r'})}{\omega - \epsilon^{\sigma}_{a} + i\eta}, \label{eq:G}
\end{equation}
where $\sigma$ is the spin direction, $\omega$ is frequency (energy), $\eta$ is a positive infinitesimal, $\epsilon^{\sigma}_{m}$ and $\varphi^{\sigma}_{m}$ are one-electron energies and wavefunctions, respectively, and $i$ and $a$ run over occupied and empty (unoccupied or virtual) states, respectively. In the following, we omit space and frequency variables ($\mathbf{r}$, $\mathbf{r'}$, $\omega$) for notational simplicity whenever needed.

Once the non-interacting (bare) Green's function $G^{\sigma}_{0}$ is known, the interacting (dressed) Green's function $G^{\sigma}$ can be obtained by solving the Dyson equation:
\begin{equation}
G^{\sigma} = G^{\sigma}_{0} + G^{\sigma}_{0} \Delta \Sigma^{\sigma} G^{\sigma}, \label{eq:G.Dyson}
\end{equation}
where $\Sigma^{\sigma}$ is the non-local, frequency-dependent (dynamical), and non-Hermitian self-energy, which accounts for many-body effects. Two popular approximations to the self-energy are the $GW$ approximation and second-order M{\o}ller--Plesset perturbation theory (MP2), which are based on $W$ and the bare (unscreened) Coulomb interaction $v$, respectively.~\cite{Ren12,Bruneval21}

\subsection{Self-consistent field (SCF)} \label{sec:SCF}

In MBPT for molecules using localized orbitals, the molecular orbitals (MOs) and MO energies are used as one-electron wavefunctions and energies, respectively, to construct the Green's function in Eq.~(\ref{eq:G}). MOs can be expressed as a linear combination of atomic orbitals $\phi_{\mu}$ (LCAO):
\begin{equation}
\varphi^{\sigma}_{m}(\mathbf{r}) = \sum_{\mu} C^{\sigma}_{\mu m} \phi_{\mu}(\mathbf{r}), \label{eq:MO}
\end{equation}
where $C^{\sigma}_{\mu m}$ are the LCAO-MO coefficients. MOLGW uses Gaussian-type orbitals (GTOs) as AOs.

The LCAO-MO coefficients in Eq.~(\ref{eq:MO}) can be obtained from a self-consistent solution to a generalized eigenvalue problem:
\begin{equation}
\mathbf{H}^{\sigma} \mathbf{C}^{\sigma} = \mathbf{S} \mathbf{C}^{\sigma} \mathbf{\epsilon}^{\sigma}, \label{eq:GEP}
\end{equation}
where $\mathbf{C}^{\sigma}$ is a matrix of LCAO-MO coefficients, $\mathbf{\epsilon}^{\sigma}$ is a diagonal matrix of MO energies, $\mathbf{S}$ is the AO overlap matrix:
\begin{equation}
S_{\mu\nu} = \int d\mathbf{r} \phi_{\mu}(\mathbf{r}) \phi_{\nu}(\mathbf{r}), \label{eq:S}
\end{equation}
and $\mathbf{H}^{\sigma}$ is the Hamiltonian matrix:
\begin{equation}
H^{\sigma}_{\mu\nu} = T_{\mu\nu} + V_{\text{ext},\mu\nu} + J_{\mu\nu} - K^{\sigma}_{\mu\nu} + \Sigma^{\sigma}_{\text{c},\mu\nu}, \label{eq:H}
\end{equation}
where $T$, $V_{\text{ext}}$, $J$, and $K^{\sigma}$ are the kinetic energy, external potential energy, Hartree, and Fock exchange terms, respectively, and $\Sigma^{\sigma}_{\text{c}}$ is the correlation part of the self-energy. When $\Sigma^{\sigma}_{\text{c}} = 0$ (neglecting electron correlation effects) in Eq.~(\ref{eq:H}), Eq.~(\ref{eq:GEP}) becomes the HF equation.

We parallelize the last three terms on the right side of Eq.~(\ref{eq:H}) -- $J$, $K^{\sigma}$, and $\Sigma^{\sigma}_{\text{c}}$ -- using OpenACC in this work. The Hartree term in Eq.~(\ref{eq:H}) is given by
\begin{equation}
J_{\mu\nu} = \sum_{\lambda\tau} (\mu\nu|\lambda\tau) \sum_{\sigma} D^{\sigma}_{\lambda\tau}, \label{eq:J}
\end{equation}
where $(\mu\nu|\lambda\tau)$ are the 4-center two-electron Coulomb repulsion integrals:
\begin{equation}
(\mu\nu|\lambda\tau) = \iint d\mathbf{r} d\mathbf{r'} \phi_{\mu}(\mathbf{r}) \phi_{\nu}(\mathbf{r}) \frac{1}{|\mathbf{r} - \mathbf{r'}|} \phi_{\lambda}(\mathbf{r'}) \phi_{\tau}(\mathbf{r'}), \label{eq:AO.4.ERI}
\end{equation}
and $\mathbf{D}^{\sigma}$ is the density matrix:
\begin{equation}
D^{\sigma}_{\mu\nu} = \sum_{m} f^{\sigma}_{m} C^{\sigma}_{\mu m} C^{\sigma}_{\nu m}, \label{eq:D}
\end{equation}
where $f^{\sigma}_{m}$ are occupation numbers (0 or 1). The Fock exchange term in Eq.~(\ref{eq:H}) is given by
\begin{equation}
K^{\sigma}_{\mu\nu} = \sum_{\lambda\tau} D^{\sigma}_{\lambda\tau} (\mu\lambda|\tau\nu). \label{eq:K}
\end{equation}

%which is beyond the scope of this work.

\textcolor{black}{It should be noted that the resolution-of-identity (RI) approximation (the density-fitting approximation) significantly reduces the computational and memory costs of the 4-center integrals in Eq.~(\ref{eq:AO.4.ERI}) at the cost of a slight accuracy loss.~\cite{Ren12,Byun19} In MOLGW, the RI approximation is parallelized using message passing interface (MPI).~\cite{Bruneval16} In this work, the MPI-parallelized RI approximation in MOLGW is not ported to the GPU, because the RI-MP2 implementation in MOLGW suffers from the MPI communication overhead (the communication time is much greater than the computation time),~\cite{Byun21} and fixing the issue is beyond the scope of this work.}

%In this work, we does not use the RI approximation, because (i) RI produces small errors, leading to accuracy loss,~\cite{Ren12,Byun19} (ii) aug-cc-pVDZ basis sets using RI are not available for Cu and Zn, and (iii) in MOLGW, MP2 implementations using RI suffer from the MPI communication overhead.~\cite{Byun21}

\subsection{$GW$ approximation} \label{sec:GW}

The $GW$ approximation neglecting vertex corrections (electron-hole interactions) is the first-order expansion of the self-energy in $W$. The interacting (reducible) polarizability $\chi$ is needed to obtain the correlation part of the $GW$ self-energy $\Sigma^{\text{GW},\sigma}_{\text{c}}$:
\begin{equation}
\Sigma^{\text{GW},\sigma}_{\text{c}} = i G^{\sigma} (W - v) = i G^{\sigma} v \chi v, \label{eq:Sigma.GW.c}
\end{equation}
and $\chi$ within the random-phase approximation (RPA) can be obtained by solving the Casida equation:
\begin{equation}
 \begin{pmatrix*}[r]
  \mathbf{A} & \mathbf{B} \\
  -\mathbf{B} & -\mathbf{A}
 \end{pmatrix*}
 \begin{pmatrix}
  X^{s} \\
  Y^{s}
 \end{pmatrix}
 =
 \begin{pmatrix}
  X^{s} \\
  Y^{s}
 \end{pmatrix}
 \Omega_{s}, \label{eq:Casida.AB.XY}
\end{equation}
where $\mathbf{A}$ and $\mathbf{B}$ are the resonant and coupling matrices, respectively, and $(X^{s}, Y^{s})$ and $\Omega_{s}$ are the eigenvectors and corresponding eigenvalues, respectively. $\mathbf{A}$ and $\mathbf{B}$ are given by
\begin{align}
A^{jb\sigma'}_{ia\sigma} &= (\epsilon^{\sigma}_{a} -\epsilon^{\sigma}_{i}) \delta_{ij} \delta_{ab} \delta_{\sigma\sigma'} \textcolor{black}{-} (ia\sigma|jb\sigma'), \label{eq:Casida.A} \\
B^{jb\sigma'}_{ia\sigma} &= \textcolor{black}{-}(ia\sigma|bj\sigma'), \label{eq:Casida.B}
\end{align}
where $i$ and $j$ are for occupied states, $a$ and $b$ are for empty states, and $(ia\sigma|jb\sigma')$ are the 4-orbital two-electron Coulomb repulsion integrals:
\begin{equation}
(ia\sigma|jb\sigma') = \iint d\mathbf{r} d\mathbf{r'} \varphi^{\sigma}_{i}(\mathbf{r}) \varphi^{\sigma}_{a}(\mathbf{r}) \frac{1}{|\mathbf{r} - \mathbf{r'}|} \varphi^{\sigma'}_{j}(\mathbf{r'}) \varphi^{\sigma'}_{b}(\mathbf{r'}), \label{eq:MO.4.ERI}
\end{equation}
which is obtained from the AO-MO integral transformation:
\begin{equation}
(ia\sigma|jb\sigma') = \sum_{\mu\nu\lambda\tau} C^{\sigma}_{\mu i} C^{\sigma}_{\nu a} C^{\sigma'}_{\lambda j} C^{\sigma'}_{\tau b} (\mu\nu|\lambda\tau). \label{eq:AO.to.MO}
\end{equation}

Once $(X^{s}, Y^{s})$ and $\Omega_{s}$ in Eq.~(\ref{eq:Casida.AB.XY}) are found, one can obtain the spectral (Lehmann) representation of $\chi({\omega})$, $W({\omega})$, and $\Sigma^{\text{GW},\sigma}_{\text{c}}(\omega)$ successively, as shown in Eq.~(\ref{eq:Sigma.GW.c}). $\Sigma^{\text{GW},\sigma}_{\text{c}}(\omega)$ is given by 
\begin{align}
\langle \varphi^{\sigma}_{m} | \Sigma^{\text{GW},\sigma}_{\text{c}} (\omega) | \varphi^{\sigma}_{n} \rangle &= \sum_{is} \frac{w^{s}_{mi\sigma} w^{s}_{ni\sigma}}{\omega - \epsilon^{\sigma}_{i} + \Omega_{s} - i\eta} \nonumber \\
&+ \sum_{as} \frac{w^{s}_{ma\sigma} w^{s}_{na\sigma}}{\omega - \epsilon^{\sigma}_{a} - \Omega_{s} + i\eta}, \label{eq:Sigma.c.GW}
\end{align}
where $i$ runs over occupied states, $a$ runs over empty states, $s$ runs over all excitations, and $w^{s}_{mn\sigma}$ are given by
\begin{equation}
w^{s}_{mn\sigma} = \sum_{ia\sigma'} (mn\sigma|ia\sigma') (X^{s}_{ia\sigma'} + Y^{s}_{ia\sigma'}). \label{eq:w.residue}
\end{equation}
The conventional $G_{0}W_{0}$ method uses only diagonal elements ($m = n$) of Eq.~(\ref{eq:Sigma.c.GW}) for efficiency, giving only quasiparticle (QP) energies.

In order to remove the starting point dependency in the $G_{0}W_{0}$ method, Faleev and co-workers proposed the qs$GW$ method using a static and Hermitian approximation to the $GW$ self-energy.~\cite{Faleev04,vanSchilfgaarde06,Kotani07} The correlation part of the ``mode A'' qs$GW$ self-energy $\Sigma^{\text{qsGW},\sigma}_{\text{c}}$ is given by
\begin{align}
&\langle \varphi^{\sigma}_{m} | \Sigma^{\text{qsGW},\sigma}_{\text{c}} | \varphi^{\sigma}_{n} \rangle \nonumber \\
&= \frac{1}{2} \left[ \langle \varphi^{\sigma}_{m} | \Sigma^{\text{GW},\sigma}_{\text{c}} (\epsilon^{\sigma}_{n}) | \varphi^{\sigma}_{n} \rangle + \langle \varphi^{\sigma}_{n} | \Sigma^{\text{GW},\sigma}_{\text{c}} (\epsilon^{\sigma}_{m}) | \varphi^{\sigma}_{m} \rangle \right], \label{eq:Sigma.c.qsGW.real}
\end{align}
where $\epsilon^{\sigma}_{m}$ and $\varphi^{\sigma}_{m}$ are qs$GW$ QP energies and wavefunctions, respectively. Eq.~(\ref{eq:Sigma.c.qsGW.real}) is referred to as the quasiparticle self-consistent appoximation in this work. When $\Sigma^{\sigma}_{\text{c}} = \Sigma^{\text{qsGW},\sigma}_{\text{c}}$ in Eq.~(\ref{eq:H}), Eq.~(\ref{eq:GEP}) becomes the qs$GW$ QP equation, which updates both eigenvalues and eigenvectors and thus gives both QP energies and wavefunctions.~\cite{Bruneval06}

\textcolor{black}{It should be noted that} for the construction of the $\Sigma^{\text{qsGW},\sigma}_{\text{c}}$ matrix \textcolor{black}{in Eq.~(\ref{eq:Sigma.c.qsGW.real})}, plane wave-based MBPT codes use truncated basis sets for efficiency, whereas Gaussian-based MOLGW uses all basis sets, because Gaussian basis sets are much more compact than plane wave ones.~\cite{Bruneval06,Kotani07}

%There are a couple of points to note related to the computational cost of the quasiparticle self-consistent appoximation in Eq.~(\ref{eq:Sigma.c.qsGW.real}). First, while $G_{0}W_{0}$ typically solves QP equations only for a couple of MOs of interest, such as frontier MOs, qs$GW$ solves QP equations for all MOs unless the so-called scissors shift is used. Second, for the construction of the $\Sigma^{\text{qsGW},\sigma}_{\text{c}}$ matrix, plane wave-based MBPT codes use truncated basis sets for efficiency, whereas Gaussian-based MOLGW uses all basis sets, because Gaussian basis sets are much more compact than plane wave ones.~\cite{Bruneval06,Kotani07}

\subsection{Second-order M{\o}ller--Plesset perturbation theory (MP2)} \label{sec:MP2}

MP2 is the second-order expansion of the self-energy in $v$ and is the simplest post-HF method. For molecules with weak screening, $v$ is close to $W$, so MP2 and $GW$ methods give similar results for the IE of atoms and molecules.~\cite{Ren12,Bruneval21} The correlation part of the MP2 self-energy $\Sigma^{\text{MP2},\sigma}_{\text{c}}(\omega)$ is given by~\cite{Ren12}
\begin{align}
&\langle \varphi^{\sigma}_{m} | \Sigma^{\text{MP2},\sigma}_{\text{c}} (\omega) | \varphi^{\sigma}_{n} \rangle \nonumber \\
&= \sum_{iap\sigma'} (mp\sigma | ia\sigma') (pn\sigma | ai\sigma') \nonumber \\
&\times \bigg [ \frac{f^{\sigma}_{p}}{\omega + \epsilon^{\sigma'}_{a} - \epsilon^{\sigma'}_{i} - \epsilon^{\sigma}_{p} - i\eta} + \frac{1 - f^{\sigma}_{p}}{\omega + \epsilon^{\sigma'}_{i} - \epsilon^{\sigma'}_{a} - \epsilon^{\sigma}_{p} + i\eta} \bigg ] \nonumber \\
&- \sum_{iap} (mp\sigma | ia\sigma) (pi\sigma | an\sigma) \nonumber \\
&\times \bigg [ \frac{f^{\sigma}_{p}}{\omega + \epsilon^{\sigma}_{a} - \epsilon^{\sigma}_{i} - \epsilon^{\sigma}_{p} - i\eta} + \frac{1 - f^{\sigma}_{p}}{\omega + \epsilon^{\sigma}_{i} - \epsilon^{\sigma}_{a} - \epsilon^{\sigma}_{p} + i\eta} \bigg ], \label{eq:Sigma.c.MP2}
\end{align}
where $i$ runs over occupied states, $a$ runs over empty states, $p$ runs over both occupied and empty states, and $\epsilon^{\sigma}_{m}$ and $\varphi^{\sigma}_{m}$ are HF energies and wavefunctions, respectively. On the right side of Eq.~(\ref{eq:Sigma.c.MP2}), the first summation is the second-order Coulomb (direct) interaction, and the second summation is the second-order exchange (SOX) interaction. Like the $G_{0}W_{0}$ method, the one-shot MP2 method starting from HF (MP2@HF) uses only diagonal elements ($m = n$) of Eq.~(\ref{eq:Sigma.c.MP2}), giving only QP energies.

The quasiparticle self-consistent MP2 (qsMP2) method is a MP2 counterpart of the qs$GW$ method. For molecules with strong correlation, qsMP2 gives more stable results than MP2@HF.~\cite{Byun19} Applying the quasiparticle self-consistent appoximation in Eq.~(\ref{eq:Sigma.c.qsGW.real}) to $\Sigma^{\text{MP2},\sigma}_{\text{c}}(\omega)$ in Eq.~(\ref{eq:Sigma.c.MP2}) gives the correlation part of the qsMP2 self-energy $\Sigma^{\text{qsMP2},\sigma}_{\text{c}}$:
\begin{align}
&\langle \varphi^{\sigma}_{m} | \Sigma^{\text{qsMP2},\sigma}_{\text{c}} | \varphi^{\sigma}_{n} \rangle \nonumber \\
&= \frac{1}{2} \left[ \langle \varphi^{\sigma}_{m} | \Sigma^{\text{MP2},\sigma}_{\text{c}} (\epsilon^{\sigma}_{n}) | \varphi^{\sigma}_{n} \rangle + \langle \varphi^{\sigma}_{n} | \Sigma^{\text{MP2},\sigma}_{\text{c}} (\epsilon^{\sigma}_{m}) | \varphi^{\sigma}_{m} \rangle \right], \label{eq:Sigma.c.qsMP2.real}
\end{align}
where $\epsilon^{\sigma}_{m}$ and $\varphi^{\sigma}_{m}$ are qsMP2 QP energies and wavefunctions, respectively. When $\Sigma^{\sigma}_{\text{c}} = \Sigma^{\text{qsMP2},\sigma}_{\text{c}}$ in Eq.~(\ref{eq:H}), Eq.~(\ref{eq:GEP}) becomes the qsMP2 QP equation.

It should be noted that the correlation part of the MP2 self-energy in Eq.~(\ref{eq:Sigma.c.MP2}) is different from the MP2 correlation energy. The MP2 self-energy is for excited-state properties, such as electron attachment and detachment energies, whereas the MP2 correlation energy is for ground-state properties, such as total energy. For atoms and molecules, the so-called $\Delta$SCF method based on the MP2 total energy gives similar vertical IEs to the MP2 self-energy method.~\cite{Ren12}

\textcolor{black}{
\subsection{Computational costs of qs$GW$ and qsMP2} \label{sec:costs}
}

%the highest occupied molecular orbital (HOMO)

\textcolor{black}{
In terms of \emph{compute time}, qs$GW$ and qsMP2 methods are computationally much more demanding than one-shot $G_{0}W_{0}$ and MP2@HF methods because of a couple of reasons. First, one-shot $G_{0}W_{0}$ and MP2@HF methods need no SCF iteration, but qs$GW$ and qsMP2 methods require dozens of SCF iterations.~\cite{Bruneval12,Byun21} Second, while one-shot $G_{0}W_{0}$ and MP2@HF methods need to calculate only diagonal ($m = n$) elements of the $\langle \varphi_{m} | \Sigma_{\text{c}} (\omega) | \varphi_{n} \rangle$ matrix in Eqs.~(\ref{eq:Sigma.c.GW}) and (\ref{eq:Sigma.c.MP2}), qs$GW$ and qsMP2 methods have to calculate all elements. Last, whereas one-shot $G_{0}W_{0}$ and MP2@HF methods need to solve QP equations for a couple of MOs of interest, such as frontier MOs, qs$GW$ and qsMP2 methods have to solve QP equations for all MOs.
}

%This will be discussed in detail in Section~\ref{sec:discussion}.

\textcolor{black}{
However, qs$GW$ and qsMP2 methods use almost the same amount of \emph{memory} as one-shot $G_{0}W_{0}$ and MP2@HF methods, and thus are ideal for GPU acceleration. Memory-hungry electronic-structure methods that demand a large amount of memory are not suitable for GPU computing, because they cannot run on the GPU with a small memory.
}

%HPC scientific applications
%benefit from

\textcolor{black}{
\subsection{Computational bottlenecks} \label{sec:bottlenecks}
}

\textcolor{black}{
Figure~\ref{fig:flowchart} depicts a flowchart of qs$GW$ and qsMP2 implementations in MOLGW. Figure~\ref{fig:flowchart} shows the computational bottlenecks, such as $J$ in Eq.~(\ref{eq:J}), $K$ in Eq.~(\ref{eq:K}), the AO-MO integral transformation in Eq.~(\ref{eq:AO.to.MO}), $\Sigma^{\text{qsGW}}_{\text{c}}$ in Eqs.~(\ref{eq:Sigma.c.GW}), (\ref{eq:w.residue}), and (\ref{eq:Sigma.c.qsGW.real}), and $\Sigma^{\text{qsMP2}}_{\text{c}}$ in Eqs.~(\ref{eq:Sigma.c.MP2}) and (\ref{eq:Sigma.c.qsMP2.real}). For the AO-MO integral transformation in Eq.~(\ref{eq:AO.to.MO}), the compute time scales as $O(N^5)$ with $N$ being the system size, because MOLGW uses a traditional implementation that reduces the $O(N^8)$ algorithm to $O(N^5)$.~\cite{ProgrammingChemistry} The AO-MO integral transformation is a major computational bottleneck common in qs$GW$ and qsMP2 calculations, as shown in Eqs.~(\ref{eq:AO.to.MO}), (\ref{eq:Sigma.c.GW}), (\ref{eq:w.residue}), and (\ref{eq:Sigma.c.MP2}). For the Casida matrix in Eqs.~(\ref{eq:Casida.AB.XY}), (\ref{eq:Casida.A}), and (\ref{eq:Casida.B}), the construction time and memory scale as $O(N^5)$ and $O(N^4)$, respectively, and the diagonalization time scales as $O(N^6)$. In this work, the construction and diagonalization of the Casida matrix are minor computational bottlenecks, because small Casida matrices are used.
}

\begin{figure*}
\begin{lstlisting}[frame=single, language=Fortran, caption={\textcolor{black}{Simplified Fortran source code of the OpenMP- and OpenACC-parallelized subroutine for the calculation of the correlation part of the qsMP2 self-energy}}, label=list:source.code]
! Program: MOLGW
! File: pt2_selfenergy.f90
! Description: Calculate the correlation part of the qsMP2 self-energy
! Author1: Fabien Bruneval (a serial CPU version using Fortran)
! Author2: Young-Moo Byun (a parallel CPU/GPU version using OpenMP/OpenACC)

subroutine pt2_selfenergy_qs(...)
 ...

#if !defined (_OPENMP) && defined (_OPENACC)
 !$acc data copyin(...) copy(...)
#endif

 do pqispin=1,nspin
   do istate=ncore_G+1,nvirtual_G-1
     ...
     
     ! OpenMP/OpenACC-parallelized AO-MO integral transformation
     call calculate_eri_4center_eigen(...)
     
     ...

#if defined (_OPENMP) && !defined (_OPENACC)
 !$OMP PARALLEL
 !$OMP DO PRIVATE(...) REDUCTION(+:...) COLLAPSE(2)
#endif
#if !defined (_OPENMP) && defined (_OPENACC)
 !$acc parallel loop independent collapse(2)
#endif

     do pstate=nsemin,nsemax
       do qstate=nsemin,nsemax
         ...
       enddo
     enddo
     
#if !defined (_OPENMP) && defined (_OPENACC)
 !$acc end parallel
#endif
#if defined (_OPENMP) && !defined (_OPENACC)
 !$OMP END DO
 !$OMP END PARALLEL
#endif

   enddo
 enddo
 
#if !defined (_OPENMP) && defined (_OPENACC)
 !$acc end data
#endif

 ...
end subroutine pt2_selfenergy_qs
\end{lstlisting}
\end{figure*}

%program hello
% ! This is a comment line
% print *, "Hello World!"
%end program

%the highest occupied molecular orbital (HOMO)

\begin{figure}
\begin{tabular}{c}
%{\includegraphics[trim=0mm 0mm 0mm 0mm, clip, width=0.48\textwidth]{figure_difference_CuO_GW}}
{\includegraphics[trim=0mm 0mm 0mm 0mm, clip, width=0.48\textwidth]{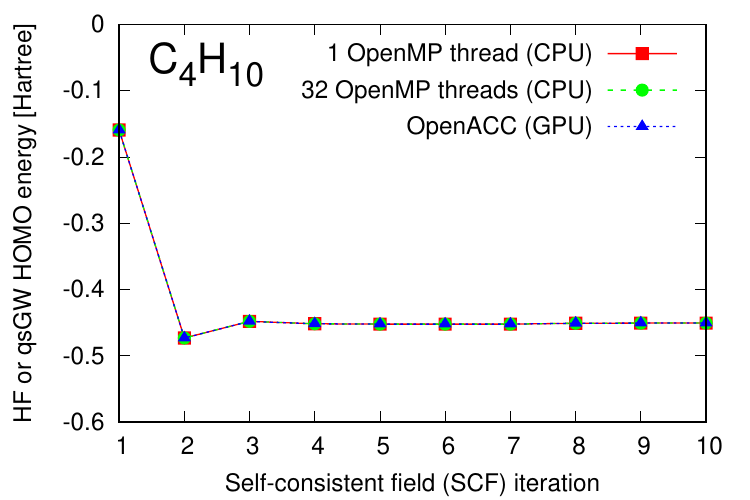}}
\end{tabular}
\caption{\textcolor{black}{(Color online) Hartree--Fock (HF) and qs$GW$ energies for the highest occupied molecular orbital (HOMO) of a butane molecule (C$_{4}$H$_{10}$) during the self-consistent field (SCF) iteration, obtained from OpenMP and OpenACC calculations. Total of 10 SCF iterations consist of first 5 HF iterations and subsequent 5 qs$GW$ iterations. The cc-pVTZ basis set is used. $\eta$ = 0.05~Hartree in Eq.~(\ref{eq:Sigma.c.GW}) is used for qs$GW$. Data is taken from Table~II in Supporting Information.}}
\label{fig:OpenMP.OpenACC.difference}
\end{figure}

\section{Implementation details} \label{sec:implementation.details}

\textcolor{black}{
\subsection{OpenACC implementation} \label{sec:OpenACC.implementation}
}

OpenMP and OpenACC are multi-platform shared-memory (thread-based) parallel programming models, which can target both CPUs and GPUs and are free from the inter-process communication overhead in distributed-memory (process-based) models such as MPI, \textcolor{black}{as discussed in Section~\ref{sec:SCF}}.~\cite{OpenMP,OpenACC,Byun21} \textcolor{black}{Although OpenMP supports GPU offloading,~\cite{Pham23} we used OpenMP for CPUs and OpenACC for GPUs in this work, because for now OpenACC is a more mature model for GPU offloading than OpenMP.}

\textcolor{black}{
Table~\ref{tab:OpenACC.directives.clauses} summarizes OpenACC directives and clauses used in this work and presents where and why they are used in the source code. Table~\ref{tab:OpenACC.directives.clauses} shows that our OpenACC implementation in MOLGW generates a total of 8 kernels for 5 computational bottlenecks and 4 kernels for the AO-MO integral transformation in Eq.~(\ref{eq:AO.to.MO}).
}

\textcolor{black}{Listing~\ref{list:source.code} presents an example of OpenMP (CPU) and OpenACC (GPU) implementations in MOLGW. Listing~\ref{list:source.code} shows that the calculation of $\Sigma^{\text{qsMP2}}_{\text{c}}$ consists of two computational bottlenecks: the AO-MO integral transformation (a line 19) and $\Sigma^{\text{qsMP2}}_{\text{c}}$ excluding the transformation (lines 31--35) (see Table~\ref{tab:OpenACC.directives.clauses}). This is the case for the calculation of $\Sigma^{\text{qsGW}}_{\text{c}}$ as well, as shown in Table~\ref{tab:OpenACC.directives.clauses}. Listing~\ref{list:source.code} also shows} that we did not mix OpenMP \textcolor{black}{(CPU)} and OpenACC \textcolor{black}{(GPU)} \textcolor{black}{(lines 10, 23, 27, 37, 40, and 48)}, making the \textcolor{black}{GPU} version of MOLGW run on a single CPU core.

We implemented OpenACC into MOLGW 1 and made our local version of MOLGW 1 with OpenMP and OpenACC implementations publicly available via GitHub.~\cite{GitHub} We will merge our OpenACC implementation into MOLGW 3, as the OpenMP implementation was merged into MOLGW 2.~\cite{Byun19}

%In this work, we used OpenMP for CPUs and OpenACC for GPUs.
%Eqs.~(\ref{eq:Sigma.c.MP2}), (\ref{eq:Sigma.c.qsMP2.real}), and (\ref{eq:AO.to.MO})
%and fair comparison purposes
%For simple implementation purposes,

%both Intel and NVIDIA compilers (formerly known as PGI compilers) for OpenMP
%dominent
%OpenMP and OpenACC implementations in MOLGW are generally similar.~\cite{Byun19} For example, both parallelize computational bottlenecks, such as $J$ in Eq.~(\ref{eq:J}), $K$ in Eq.~(\ref{eq:K}), the AO-MO integral transformation in Eq.~(\ref{eq:AO.to.MO}), $\Sigma^{\text{qsGW},\sigma}_{\text{c}}$ in Eqs.~(\ref{eq:Sigma.c.GW}), (\ref{eq:w.residue}), and (\ref{eq:Sigma.c.qsGW.real}), and $\Sigma^{\text{qsMP2},\sigma}_{\text{c}}$ in Eqs.~(\ref{eq:Sigma.c.MP2}) and (\ref{eq:Sigma.c.qsMP2.real}), using similar compiler directives. Among all bottlenecks, building $\Sigma^{\text{qsGW},\sigma}_{\text{c}}$ and $\Sigma^{\text{qsMP2},\sigma}_{\text{c}}$ matrices is the most dominant bottleneck. Also, nested loops in OpenMP and OpenACC parallel regions are collapsed (merged or unrolled) to increase parallelism in an identical fashion. 
%in an identical fashion

%\textcolor{black}
%{
%OpenMP and OpenACC implementations in MOLGW annotate the original source code with pragmas without data layout restructuring~\cite{Bonati15} and code refactoring. 
%}

\textcolor{black}{
\subsection{Similarities between OpenMP and OpenACC implementations} \label{sec:similarity.OpenMP.OpenACC}
}

\textcolor{black}{
OpenMP (CPU) and OpenACC (GPU) implementations in MOLGW are similar in a few ways.~\cite{Byun19} First, both annotate the original source code with pragmas without data layout restructuring,~\cite{Bonati15} allowing us to preserve the loop order in a serial CPU version, as shown at lines 14, 15, 31, and 32 in Listing~\ref{list:source.code}.
}

%the calculation of $\Sigma^{\text{qsGW}}_{\text{c}}$
%perform large-scale calculations
%large-scale calculations

\textcolor{black}{
Second, both require code refactoring.~\cite{CodeRefactoring} For example, we fixed a race condition, which we encountered when parallelizing the implementation of $\Sigma^{\text{qsGW}}_{\text{c}}$. Also, we changed the index type of some arrays from a 4-byte integer to an 8-byte integer to enable MOLGW to use a large amount of memory. Also, we implemented our own timing routines into MOLGW (for Section~\ref{sec:partial.timings}), because built-in timing routines in MOLGW give wrong results for long calculations.
}

Third, both parallelize the computational bottlenecks using similar compiler directives. \textcolor{black}{For example, Listing~\ref{list:source.code} shows that OpenMP and OpenACC parallelized a loop using \texttt{omp parallel do} (lines 24, 25, 41, and 42) and \texttt{acc parallel loop} (lines 28 and 38) directives, respectively.}

Last, both collapse (merge or unroll) nested loops into a single loop to increase parallelism. \textcolor{black}{For example, Listing~\ref{list:source.code} shows that OpenMP and OpenACC merged \texttt{pstate} and \texttt{qstate} loops (lines 31, 32, 34, and 35) using the \texttt{collapse} clause (lines 25 and 28, respectively).}

%in parallel regions
%such as $J$ in Eq.~(\ref{eq:J}), $K$ in Eq.~(\ref{eq:K}), the AO-MO integral transformation in Eq.~(\ref{eq:AO.to.MO}), $\Sigma^{\text{qsGW},\sigma}_{\text{c}}$ in Eqs.~(\ref{eq:Sigma.c.GW}), (\ref{eq:w.residue}), and (\ref{eq:Sigma.c.qsGW.real}), and $\Sigma^{\text{qsMP2},\sigma}_{\text{c}}$ in Eqs.~(\ref{eq:Sigma.c.MP2}) and (\ref{eq:Sigma.c.qsMP2.real})
% Among all bottlenecks, building $\Sigma^{\text{qsGW},\sigma}_{\text{c}}$ and $\Sigma^{\text{qsMP2},\sigma}_{\text{c}}$ matrices is the most dominant bottleneck.
%NVIDIA profiling tools
%to build the $\Sigma^{\text{GW},\sigma}_{\text{c}}$ matrix in Eqs.~(\ref{eq:Sigma.c.GW}) and (\ref{eq:w.residue}).
%,\sigma
%as shown in Table~\ref{tab:OpenACC.directives.clauses}
%for CPUs
%for GPUs
%a single instruction multiple data (SIMD)-vectorized CPU library
%making a small effect on the performance comparison of OpenMP and OpenACC implementations.
%the Libint library
%because it takes negligible time to diagonalize small matrices used in this work.
%,\sigma
%Fig.~\ref{fig:flowchart} and
%using single instruction multiple data (SIMD) vectorization

\textcolor{black}{
\subsection{Differences between OpenMP and OpenACC implementations} \label{sec:difference.OpenMP.OpenACC}
}

However, there are a few differences between OpenMP \textcolor{black}{(CPU)} and OpenACC \textcolor{black}{(GPU)} implementations in MOLGW. First, unlike OpenMP, OpenACC decouples data movement from computation, \textcolor{black}{as shown at lines 11 and 49 in Listing~\ref{list:source.code}}, requiring to optimize the data transfer between host (CPU) and device (GPU) memories after offloading the computation to the GPU. We used CUDA unified (managed) memory, a single memory space for both host and device memories,~\cite{UniMem} to manage dynamically allocated arrays used \textcolor{black}{in the implementation of $\Sigma^{\text{qsGW}}_{\text{c}}$ (see Table~\ref{tab:OpenACC.directives.clauses})}. We used a manual deep copy to handle dynamic data structures, as is done for the OpenACC version of VASP,~\cite{Maintz18b} because OpenACC 2.6 does not support a true deep copy \textcolor{black}{(NVIDIA compilers used in this work support the OpenACC 2.6 specification)}. \textcolor{black}{We transfered global data from CPU memory to GPU memory after the completion of the calculation of the 4-center two-electron Coulomb repulsion integrals in Eq.~(\ref{eq:AO.4.ERI}) (see Table~\ref{tab:OpenACC.directives.clauses}).} We used NVIDIA \textcolor{black}{profilers} to confirm that we fully optimized the CPU--GPU communication overhead in our OpenACC implementation.

Second, unlike the OpenMP \textcolor{black}{(CPU)} implementation, our OpenACC \textcolor{black}{(GPU)} implementation does not parallelize the 4-center integrals in Eq.~(\ref{eq:AO.4.ERI}), because MOLGW uses Libint, a single instruction multiple data (SIMD)-vectorized CPU library for computing Gaussian integrals.~\cite{Valeev16} \textcolor{black}{The 4-center integrals are calculated only once at the beginning of MOLGW execution, and thus are a major computational bottleneck in non-self-consistent $G_{0}W_{0}$ and MP2@HF calculations, but a minor computational bottleneck in quasiparticle self-consistent $GW$ and MP2 calculations (see Section~\ref{sec:costs}).}

Last, the OpenMP \textcolor{black}{(CPU)} implementation diagonalizes matrices, such as HF, qs$GW$, and qsMP2 matrices in Eqs.~(\ref{eq:GEP}) and (\ref{eq:H}) with $\Sigma^{\sigma}_{\text{c}}$ = 0, $\Sigma^{\text{qsGW}}_{\text{c}}$, and $\Sigma^{\text{qsMP2}}_{\text{c}}$, respectively, and the Casida matrix in Eq.~(\ref{eq:Casida.AB.XY}) (\textcolor{black}{see Fig.~\ref{fig:flowchart}}), using OpenMP-parallelized Intel math kernel library (MKL), but our OpenACC \textcolor{black}{(GPU)} implementation performs the matrix diagonalization serially. \textcolor{black}{In this work, the matrix diagonalization is a minor computational bottleneck, because small matrices are used, as discussed in Section~\ref{sec:bottlenecks}.}

\textcolor{black}{
\subsection{Regression testing} \label{sec:regression.testing}
}

%the highest occupied molecular orbital (HOMO)
%shows HF and qs$GW$ energies

In order to ensure that our parallel GPU version of MOLGW gives the same results as the original serial CPU version, we performed two kinds of regression \textcolor{black}{testing}.~\cite{RegressionTesting} First, we used the built-in automated regression test suite in MOLGW and verified that our GPU implementation produces correct results. Second, we compared QP energies obtained from CPU and GPU calculations. Figure~\ref{fig:OpenMP.OpenACC.difference} \textcolor{black}{depicts} HF and qs$GW$ energies for the highest occupied molecular orbital (HOMO) of a butane molecule (C$_{4}$H$_{10}$) from serial CPU, parallel OpenMP CPU, and parallel OpenACC GPU calculations during the SCF iteration, \textcolor{black}{which are presented in Table~II in Supporting Information. Figure~\ref{fig:OpenMP.OpenACC.difference} shows} that parallel OpenMP CPU and OpenACC GPU calculations give identical results to the serial CPU calculation, as is the case for qsMP2 as well (see Fig.~1 in Supporting Information). \textcolor{black}{Table~II in Supporting Information shows 5 and and 10 matching significant digits for HOMO and total energies, respectively, during 10 SCF iterations.}

%QP
%three kinds of qs$GW$ QP energies
%CuO$^{-}$

\begin{table*}
	  \caption{\textcolor{black}{Specifications for GPUs and CPUs used in this work. GB, (G)DDR, FP64, and GFLOP represent gigabyte, (graphics) double data rate, double-precision floating-point performance, and giga floating-point operations, respectively.}}
  \label{tab:GPU.CPU.specs}
  \begin{tabular*}{1.00\textwidth}{ @{\extracolsep{\fill}} l c c c c}
    \hline \hline
& \multicolumn{2}{c}{\textcolor{black}{GPU}} & \multicolumn{2}{c}{\textcolor{black}{CPU}} \\
\cline{2-3}
\cline{4-5}
& RTX 3090\footnotemark[1] & RTX 4090\footnotemark[2] & \textcolor{black}{Xeon Phi}\footnotemark[3] & \textcolor{black}{Threadripper}\footnotemark[4] \\
%& NVIDIA GeForce RTX 3090 & NVIDIA GeForce RTX 4090 & Intel Xeon Phi 7250 & AMD Ryzen Threadripper PRO 3975WX \\
\hline
\textcolor{black}{Platform} & \textcolor{black}{Desktop} & \textcolor{black}{Desktop} & \textcolor{black}{Supercomputer} & \textcolor{black}{Workstation} \\
Release year & 2020 & 2022 & \textcolor{black}{2016} & \textcolor{black}{2020} \\
Architecture & Ampere & Ada Lovelace & \textcolor{black}{Knights Landing (KNL)} & \textcolor{black}{Zen 2 (Castle Peak)} \\
\textcolor{black}{Manufacturing process} & \textcolor{black}{8 nm} & \textcolor{black}{5 nm} & \textcolor{black}{14 nm} & \textcolor{black}{7 nm} \\
Number of cores & 10496\footnotemark[5] & 16384\footnotemark[5] & \textcolor{black}{68} & \textcolor{black}{32} \\
Boost clock & 1.70 GHz & 2.52 GHz & \textcolor{black}{1.60 GHz} & \textcolor{black}{4.20 GHz} \\
Base clock & 1.40 GHz & 2.23 GHz & \textcolor{black}{1.40 GHz} & \textcolor{black}{3.50 GHz} \\
Memory size & 24 GB & 24 GB & \textcolor{black}{96 GB} & \textcolor{black}{128 GB} \\
Memory type & GDDR6X & GDDR6X & \textcolor{black}{DDR4} & \textcolor{black}{DDR4} \\
Memory bandwidth & 935.8 GB/s & 1008 GB/s & \textcolor{black}{115.2 GB/s} & \textcolor{black}{102.4 GB/s}\footnotemark[6] \\
FP64 & 556 GFLOP/s & 1290 GFLOP/s & \textcolor{black}{N/A}\footnotemark[7] & \textcolor{black}{268 GFLOP/s}\footnotemark[8] \\
%& RTX 2080 SUPER & RTX 3090 \\
%\hline
%Release year & 2019 & 2020 \\
%Architecture & Turing & Ampere \\
%%CUDA cores & 3072 & 10496 \\
%Memory size & 8 GB & 24 GB \\
%Memory type & GDDR6 & GDDR6X \\
%Memory bandwidth & 496 GB/s & 936 GB/s \\
%FP64 & 349 GFLOPS & 556 GFLOPS \\
    \hline \hline
  \end{tabular*}
  \footnotetext[1]{NVIDIA GeForce RTX 3090}
  \footnotetext[2]{NVIDIA GeForce RTX 4090}
  \footnotetext[3]{Intel Xeon Phi 7250}
  \footnotetext[4]{AMD Ryzen Threadripper PRO 3975WX}
  \footnotetext[5]{Number of compute unified device architecture (CUDA) cores}
  \footnotetext[6]{Maximum memory bandwidth for the quad-channel configuration used in this work. The Intel Advisor gives a measured value of 61.2~GB/s.}
  \footnotetext[7]{Not available. We can no longer access the Xeon Phi to measure its FP64 using the Intel Advisor, because our supercomputing time grant at the KISTI National Supercomputing Center ended.}
  \footnotetext[8]{Measured by the Intel Advisor}
\end{table*}

%CUDA cores
%Specifications for NVIDIA GeForce RTX 2080 SUPER and RTX 3090 GPUs. GB, GDDR, FP64, and GFLOPS represent gigabyte, graphics double data rate, double-precision floating-point performance, and giga floating operations per second, respectively.
%Specifications for NVIDIA GeForce RTX 3090 and 4090 GPUs. CUDA, GB, GDDR, FP64, and GFLOPS represent compute unified device architecture, gigabyte, graphics double data rate, double-precision floating-point performance, and giga floating operations per second, respectively.
%evaluated
%about 10 times more
%We \textcolor{black}{analyzed} the performance of our GPU acceleration in MOLGW
%in Section~\ref{sec:implementation.details}

\section{Benchmark configurations} \label{sec:benchmark.configurations}

\textcolor{black}{In order to assess the performance of our OpenACC implementation in MOLGW for GPU computing, we conducted a few benchmarks} using various software and hardware configurations. We used the following software configurations: First, we used two different starting-point-independent MBPT methods, qs$GW$ and qsMP2, which are computationally the most demanding electronic structure methods in MOLGW. Second, we used the Dunning's correlation-consistent basis set, cc-pVTZ. Third, we used the first seven linear alkanes (C$_{n}$H$_{2n+2}$, where $n$ = 1, 2, ..., 7) to systemically increase the molecular size and the GPU memory usage. \textcolor{black}{We did not use alkanes larger than heptane (C$_{7}$H$_{16}$) due to our GPU memory limit of 24~GB. To calculate large alkanes, one should reduce the memory usage by using small basis sets, such as cc-pVDZ and Pople's basis sets, at the cost of accuracy loss or by implementing molecular symmetry into MOLGW.} \textcolor{black}{Fourth, for the performance analysis using compute times,} we used total of 8 SCF iterations, consisting of first 5 HF iterations [$\Sigma_{\text{c}} = 0$ in Eq.~(\ref{eq:H})] and subsequent 3 qs$GW$ or qsMP2 iterations [$\Sigma_{\text{c}} = \Sigma^{\text{qsGW}}_{\text{c}}$ or $\Sigma^{\text{qsMP2}}_{\text{c}}$, respectively, in Eq.~(\ref{eq:H})] \textcolor{black}{(see Fig.~\ref{fig:flowchart})}. It should be noted that \textcolor{black}{dozens of} qs$GW$ or qsMP2 iterations are needed to reach the convergence.~\cite{Bruneval12,Byun21} \textcolor{black}{Fifth, for the roofline performance analysis,~\cite{Williams09} we used total of 6 SCF iterations (1 qs$GW$ or qsMP2 iteration), the Intel Advisor (OpenMP), and the NVIDIA Nsight Compute (OpenACC). Last,} we used either Intel or NVIDIA compilers (formerly known as PGI compilers) for OpenMP, but only NVIDIA compilers for OpenACC.

%without sacrificing accuracy
%NVIDIA compilers that we used support the OpenACC 2.6 specification.

%the system size
%^{\sigma}
%^{\sigma}
%,\sigma
%,\sigma

%We evaluated the performance of our GPU acceleration in MOLGW using various software and hardware configurations. We used the following software configurations: First, we used two different starting-point-independent MBPT methods, qs$GW$ and qsMP2, which are computationally the most demanding electronic structure methods in MOLGW. Second, we used two different molecular systems, singlet closed-shell CuO$^{-}$ and doublet open-shell ZnO$^{-}$ anions. Third, we used Dunning's augmented correlation-consistent basis sets, aug-cc-pV$n$Z ($n$ = D, T, Q, and 5), to systemically increase the GPU memory usage. Increasing the basis size has the same effect on the GPU memory usage as increasing the system size. Last, we used total of 20 SCF iterations, consisting of first 5 HF iterations [$\Sigma^{\sigma}_{\text{c}} = 0$ in Eq.~(\ref{eq:H})] and subsequent 15 qs$GW$ or qsMP2 iterations [$\Sigma^{\sigma}_{\text{c}} = \Sigma^{\text{qsGW},\sigma}_{\text{c}}$ or $\Sigma^{\text{qsMP2},\sigma}_{\text{c}}$, respectively, in Eq.~(\ref{eq:H})].

\textcolor{black}{We used the following hardware configurations: For OpenACC, we used two different GPUs, NVIDIA GeForce RTX 3090 and 4090 for a desktop computer (RTX 3090 and RTX 4090, respectively, in the following), whose specifications are summarized in Table~\ref{tab:GPU.CPU.specs}. The RTX 4090 has a newer architecture and higher FP64 performance than the RTX 3090. For OpenMP, we used two different CPUs, AMD Ryzen Threadripper PRO 3975WX with 32 heavyweight cores for a workstation and Intel Xeon Phi 7250 with 68 lightweight cores for a supercomputer (Threadripper and Xeon Phi, respectively, in the following), \textcolor{black}{whose specifications are summarized in Table~\ref{tab:GPU.CPU.specs}}. In both Threadripper and Xeon Phi, simultaneous multithreading (SMT)\textcolor{black}{, which Intel calls hyper-threading,} is disabled. \textcolor{black}{The effect of SMT on the OpenMP (CPU) performance is presented in Table~I in Supporting Information.}}

%as is commonly done in HPC supercomputing centers due to security reasons.

%First, we used a PC with a low-end desktop GPU instead of a workstation or supercomputer with a high-end data-center GPU, such as NVIDIA V100 and A100, to make it easy for others to reproduce our benchmark results.

%~\cite{RTX3080}

%We used the following hardware configurations: First, we used a PC with a low-end desktop GPU instead of a workstation or supercomputer with a high-end data-center GPU, such as NVIDIA V100 and A100, to make it easy for others to reproduce our benchmark results. Second, we used two different GPUs, NVIDIA GeForce RTX 2080 SUPER and RTX 3090,~\cite{RTX3080} whose specifications are summarized in Table~\ref{tab:GPU.CPU.specs}. RTX 3090 has a newer architecture, larger and faster GPU memory, and higher FP64 performance than RTX 2080 SUPER. Last, we used 16 OpenMP threads on the AMD Ryzen 9 3900X CPU with simultaneous multithreading (SMT) enabled, because there are many desktop and laptop CPUs with 16 threads available these days.

\begin{table*}
  \caption{\textcolor{black}{OpenMP and OpenACC compute times using different alkane molecules (C$_{n}$H$_{2n+2}$), MBPT methods, CPUs, and GPUs. GB represents gigabyte.}}
  \label{tab:OpenMP.OpenACC.times}
%  \begin{tabular}{ c c c c c c c c c c c c }
  \begin{tabular*}{1.00\textwidth}{ @{\extracolsep{\fill}} l l r r r r r r r r r }
    \hline \hline
%System & Method & Basis set & NBF & Peak memory & OpenMP (1T) & OpenMP (16T) & OpenACC (2080) & OpenACC (3090) \\
& & & & \multicolumn{3}{c}{OpenMP \textcolor{black}{(hours)}} & \multicolumn{2}{c}{OpenACC \textcolor{black}{(hours)}} & \multicolumn{2}{c}{Speedup} \\
\cline{5-7}
\cline{8-9}
\cline{10-11}
& & & & \multicolumn{2}{c}{Threadripper\footnotemark[1]\footnotemark[3]\footnotemark[10]} & \multicolumn{1}{c}{Xeon Phi\footnotemark[2]\footnotemark[3]\footnotemark[4]\footnotemark[11]} & & & Threadripper\footnotemark[12] & Xeon Phi\footnotemark[13] \\
\cline{5-6}
\cline{7-7}
\cline{10-10}
\cline{11-11}
Molecule & Method\footnotemark[5] & NBF\footnotemark[6] & Memory\footnotemark[7] & 1 thread & 32 threads & 68 threads & RTX 3090\footnotemark[8]\footnotemark[10] & RTX 4090\footnotemark[9]\footnotemark[10] & RTX 4090\footnotemark[12] & RTX 4090\footnotemark[13] \\
\hline
CH$_{4}$              & qs$GW$ &   86 &   0.1 GB & \textcolor{black}{0.017} & \textcolor{black}{0.0014} & 0.0047 & \textcolor{black}{0.0061} & \textcolor{black}{0.0031} & \textcolor{black}{0.5x} & \textcolor{black}{1.5x} \\
C$_{2}$H$_{6}$   & qs$GW$ & 144 &   0.8 GB &   \textcolor{black}{0.33} &   \textcolor{black}{0.018} &   0.038 &   \textcolor{black}{0.043} &   \textcolor{black}{0.018} & \textcolor{black}{1.0x} & \textcolor{black}{2.1x} \\
C$_{3}$H$_{8}$   & qs$GW$ & 202 &   2.8 GB &   \textcolor{black}{2.47} &     \textcolor{black}{0.15} &     0.27 &     \textcolor{black}{0.17} &   \textcolor{black}{0.077} & \textcolor{black}{1.9x} & \textcolor{black}{3.5x} \\
C$_{4}$H$_{10}$ & qs$GW$ & 260 &   5.0 GB & \textcolor{black}{11.38} &     \textcolor{black}{0.65} &     1.00 &     \textcolor{black}{0.55} &     \textcolor{black}{0.29} & \textcolor{black}{2.2x} & \textcolor{black}{3.4x} \\
C$_{5}$H$_{12}$ & qs$GW$ & 318 &   9.5 GB &    \textcolor{black}{N/A}\footnotemark[14] &     \textcolor{black}{2.65} &     3.31 &     \textcolor{black}{1.39} &     \textcolor{black}{0.87} & \textcolor{black}{3.0x} & \textcolor{black}{3.8x} \\
C$_{6}$H$_{14}$ & qs$GW$ & 376 & 14.9 GB &    \textcolor{black}{N/A}\footnotemark[14] &     \textcolor{black}{6.93} &     9.44 &     \textcolor{black}{3.03} &     \textcolor{black}{2.17} & \textcolor{black}{3.2x} & \textcolor{black}{4.4x} \\
C$_{7}$H$_{16}$ & qs$GW$ & 434 & 21.6 GB\footnotemark[15] &    \textcolor{black}{N/A}\footnotemark[14] &   \textcolor{black}{15.69} &   23.71 &     \textcolor{black}{6.80} &     \textcolor{black}{4.98} & \textcolor{black}{3.2x} & \textcolor{black}{4.8x} \\
\hline
CH$_{4}$              & qsMP2 &   86 &   0.1 GB &   \textcolor{black}{0.11} & \textcolor{black}{0.0056} & 0.012 & \textcolor{black}{0.0078} & \textcolor{black}{0.0039} &  \textcolor{black}{1.4x} & \textcolor{black}{3.1x} \\
C$_{2}$H$_{6}$   & qsMP2 & 144 &   0.5 GB &   \textcolor{black}{1.69} &   \textcolor{black}{0.081} &   0.16 &   \textcolor{black}{0.056} &   \textcolor{black}{0.029} &  \textcolor{black}{2.8x} & \textcolor{black}{5.5x} \\
C$_{3}$H$_{8}$   & qsMP2 & 202 &   1.8 GB &   \textcolor{black}{9.42} &     \textcolor{black}{0.44} &   0.92 &     \textcolor{black}{0.26} &     \textcolor{black}{0.11} &  \textcolor{black}{4.0x} & \textcolor{black}{8.4x} \\
C$_{4}$H$_{10}$ & qsMP2 & 260 &   4.4 GB & \textcolor{black}{31.48} &     \textcolor{black}{1.69} &   3.46 &     \textcolor{black}{0.82} &     \textcolor{black}{0.34} &  \textcolor{black}{5.0x} & \textcolor{black}{10.2x} \\
C$_{5}$H$_{12}$ & qsMP2 & 318 &   8.6 GB &    \textcolor{black}{N/A}\footnotemark[14] &     \textcolor{black}{6.06} &   9.09 &     \textcolor{black}{1.94} &     \textcolor{black}{0.81} &  \textcolor{black}{7.5x} & \textcolor{black}{11.2x} \\
C$_{6}$H$_{14}$ & qsMP2 & 376 & 14.5 GB &    \textcolor{black}{N/A}\footnotemark[14] &   \textcolor{black}{14.96} & 21.94 &     \textcolor{black}{4.12} &     \textcolor{black}{1.68} &  \textcolor{black}{8.9x} & \textcolor{black}{13.1x} \\
C$_{7}$H$_{16}$ & qsMP2 & 434 & 21.4 GB\footnotemark[15] &    \textcolor{black}{N/A}\footnotemark[14] &   \textcolor{black}{29.88} & 45.40 &     \textcolor{black}{7.89} &     \textcolor{black}{3.08} &  \textcolor{black}{9.7x} & \textcolor{black}{14.7x} \\
    \hline \hline
  \end{tabular*}
  \footnotetext[1]{AMD Ryzen Threadripper PRO 3975WX 32-core CPU}
  \footnotetext[2]{Intel Xeon Phi 7250 68-core CPU}
  \footnotetext[3]{Simultaneous multithreading (SMT) is disabled.}
  \footnotetext[4]{\textcolor{black}{Single-thread calculations are not performed on the Xeon Phi CPU, which is made up of lightweight cores with a base clock of 1.40~GHz, as shown in Table~\ref{tab:GPU.CPU.specs}.}}
  \footnotetext[5]{Only 3 qs$GW$ and qsMP2 iterations are used for benchmark purposes. For real calculations, one should use about 10 times more iterations, leading to about 10 times larger OpenMP and OpenACC compute times.}
  \footnotetext[6]{\textcolor{black}{NBF represents the number of basis functions. The cc-pVTZ basis set is used.}}
  \footnotetext[7]{GPU peak memory usage}
  \footnotetext[8]{NVIDIA GeForce RTX 3090 GPU on a desktop computer with the AMD Ryzen 9 3950X 16-core CPU}
  \footnotetext[9]{NVIDIA GeForce RTX 4090 GPU on a desktop computer with the AMD Ryzen 9 5950X 16-core CPU }
  \footnotetext[10]{NVIDIA compilers (formerly known as PGI compilers) are used.}
  \footnotetext[11]{Intel compilers are used.}
  \footnotetext[12]{OpenMP time from the Threadripper CPU using 32 threads over OpenACC time from the RTX 4090 GPU}
  \footnotetext[13]{OpenMP time from the Xeon Phi CPU using 68 threads over OpenACC time from the RTX 4090 GPU}
  \footnotetext[14]{Not available because single-thread calculations are prohibitively expensive.}
  \footnotetext[15]{\textcolor{black}{qs$GW$ and qsMP2 calculations of alkanes larger than C$_{7}$H$_{16}$ are not performed on GPUs, because those calculations require more memory than available on the RTX 3090 and 4090 GPUs (24~GB, as shown in Table~\ref{tab:GPU.CPU.specs}).}}
\end{table*}

%of large molecules
%alkane molecules (C$_{n}$H$_{2n+2}$) with $n \ge 8$
%with 10,496 cores (see Table~\ref{tab:GPU.CPU.specs})
%with 16,384 cores (see Table~\ref{tab:GPU.CPU.specs})
%OpenMP and OpenACC compute times (in hours) using different molecular systems, MBPT methods, basis sets, number of CPU threads, and GPUs. NVIDIA compilers (formerly known as PGI compilers) are used for both OpenMP and OpenACC. NBF and GB represent the number of basis functions and the gigabyte, respectively.
%System
%\footnotetext[1]{OpenMP and OpenACC versions of MOLGW consume almost the same amount of CPU and GPU memories, respectively.}
%OpenMP CPU
%Speedup\footnotemark[4]
%\footnotetext[11]{OpenMP time with 16 CPU threads over OpenACC time with the NVIDIA GeForce RTX 3090 GPU}
%NVIDIA compilers (formerly known as PGI compilers) are used for both OpenMP and OpenACC.
%\footnotetext[12]{Not available because aug-cc-pV5Z calculations with a single OpenMP CPU thread are prohibitively expensive.}
%\footnotetext[12]{Not available because aug-cc-pV5Z calculations require more memory than available on the NVIDIA GeForce RTX 2080 SUPER GPU (8~GB, as shown in Table~\ref{tab:GPU.CPU.specs}).}
%NBF and GB represent the number of basis functions and the gigabyte, respectively.
%single-thread calculations for large molecules

\begin{figure}
\begin{tabular}{c}
{\includegraphics[trim=0mm 0mm 0mm 0mm, clip, width=0.48\textwidth]{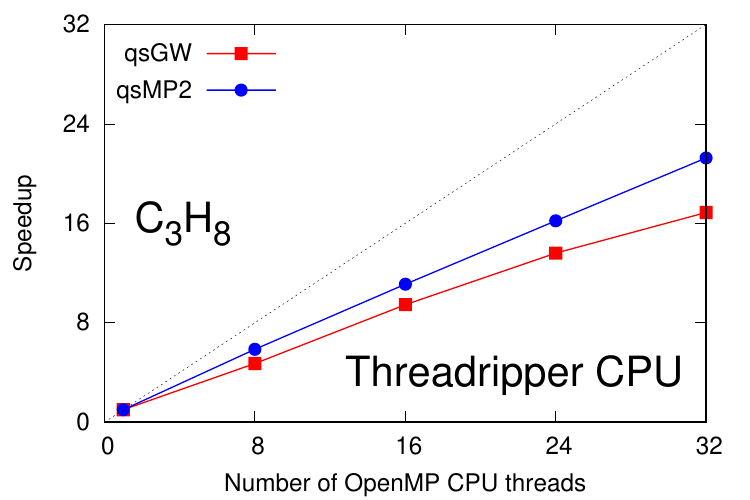}}
\end{tabular}
\caption{\textcolor{black}{(Color online) OpenMP speedup as a function of number of CPU threads using different MBPT methods. The cc-pVTZ basis set, the propane molecule (C$_{3}$H$_{8}$), and the AMD Ryzen Threadripper PRO 3975WX 32-core CPU with SMT disabled are used. \emph{GPUs are not used.} The black dotted line represents the ideal speedup. Data is taken from Table~III in Supporting Information.}}
\label{fig:OpenMP.speedup}
\end{figure}

% and compilers
%OpenMP speedup as a function of number of CPU threads using different MBPT methods. The singlet closed-shell CuO$^{-}$ anion, the aug-cc-pVTZ basis set, and the AMD Ryzen 9 3900X CPU with SMT enabled are used. GPUs are not used. The black dotted line represents the ideal speedup.
%for OpenMP

\begin{figure*}
\begin{tabular}{c}
%{\includegraphics[trim=0mm 0mm 0mm 0mm, clip, width=0.48\textwidth]{figure_openmp_openacc_speedup_CuO_gw}}
%{\includegraphics[trim=0mm 0mm 0mm 0mm, clip, width=0.48\textwidth]{figure_openmp_openacc_speedup_CuO_pt2}}
{\includegraphics[trim=0mm 0mm 0mm 0mm, clip, width=0.48\textwidth]{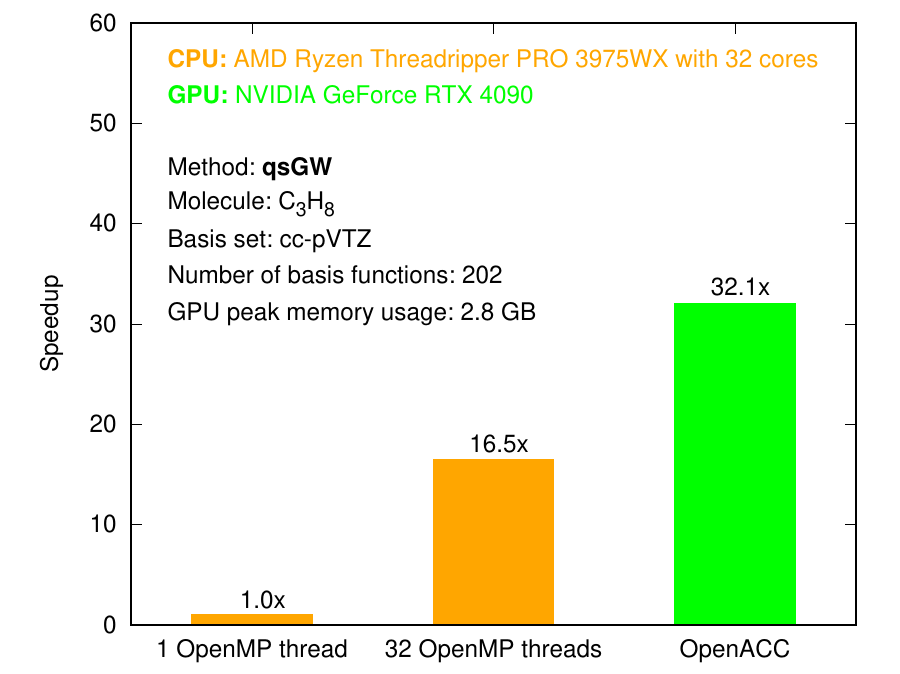}}
{\includegraphics[trim=0mm 0mm 0mm 0mm, clip, width=0.48\textwidth]{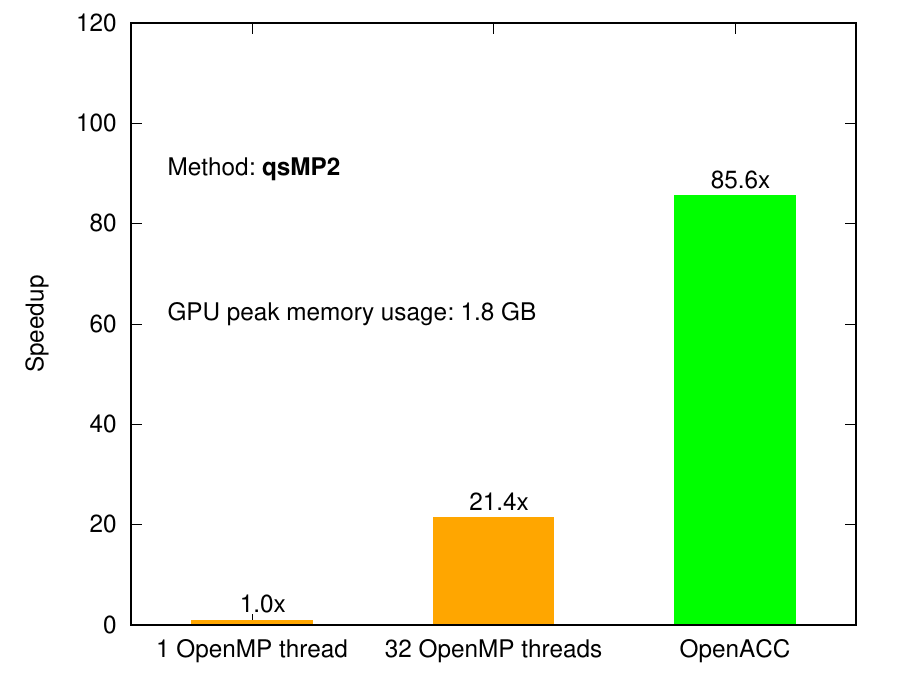}}
\end{tabular}
\caption{\textcolor{black}{(Color online) OpenMP and OpenACC speedups of qs$GW$ and qsMP2 methods (left and right, respectively). GB represents gigabyte. Data is taken from Table~\ref{tab:OpenMP.OpenACC.times}.}}
\label{fig:OpenMP.OpenACC.speedup}
\end{figure*}

%OpenMP and OpenACC speedups of qs$GW$ and qsMP2 methods (left and right, respectively). The singlet closed-shell CuO$^{-}$ anion, the aug-cc-pVQZ basis set, the AMD Ryzen 9 3900X CPU with SMT enabled, and the NVIDIA RTX 3090 GPU are used. Data is taken from Table~\ref{tab:OpenMP.OpenACC.times}.
%The propane molecule (C$_{3}$H$_{8}$) and the cc-pVTZ basis set are used. The AMD Ryzen Threadripper PRO 3975WX 32-core CPU with SMT disabled and the NVIDIA GeForce RTX 4090 GPU are used for OpenMP and OpenACC, respectively.

\begin{figure*}
\begin{tabular}{c c}
%{\includegraphics[trim=0mm 0mm 0mm 0mm, clip, width=0.48\textwidth]{figure_openacc_speedup_CuO}}
{\includegraphics[trim=0mm 0mm 0mm 0mm, clip, width=0.48\textwidth]{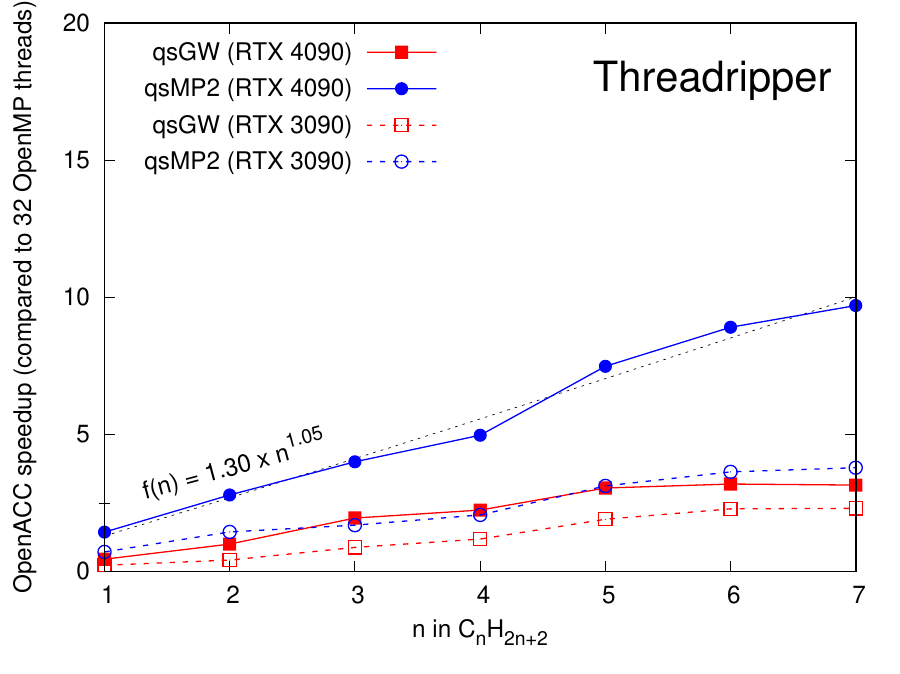}}
{\includegraphics[trim=0mm 0mm 0mm 0mm, clip, width=0.48\textwidth]{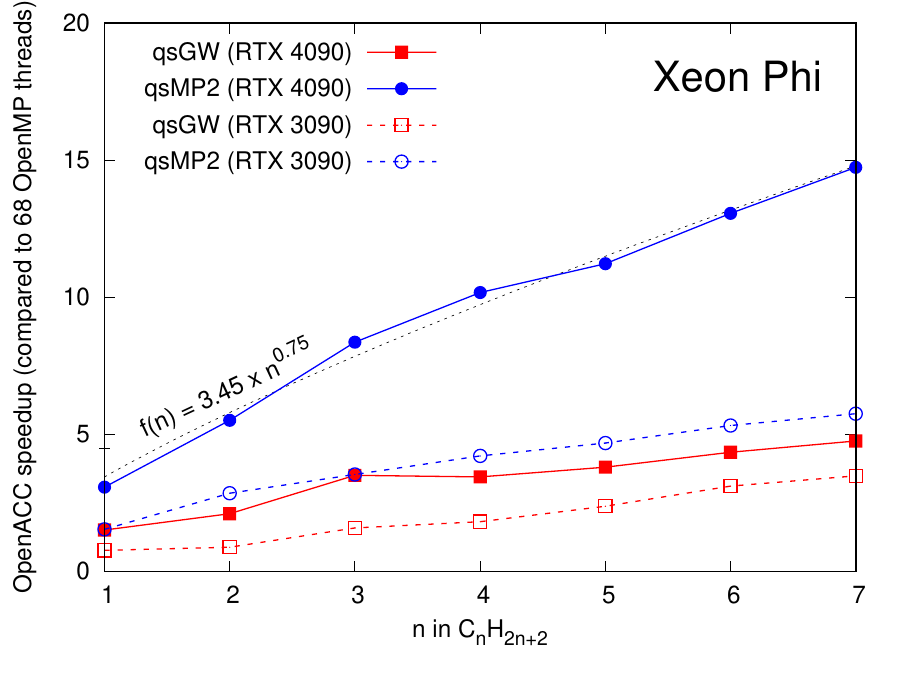}}
\end{tabular}
\caption{\textcolor{black}{(Color online) OpenACC GPU speedup relative to OpenMP CPU threads as a function of alkane (C$_{n}$H$_{2n+2}$) size using different MBPT methods and GPUs. The Threadripper 32-core and Xeon Phi 68-core CPUs (left and right, respectively) with SMT disabled are used. The cc-pVTZ basis set is used. The black dotted line is a fitted line for the qsMP2 method on the RTX 4090 GPU. Data is taken from Table~\ref{tab:OpenMP.OpenACC.times}.}}
\label{fig:OpenACC.speedup}
\end{figure*}

%on the Threadripper and Xeon Phi CPUs (left and right, respectively) with SMT disabled
%OpenACC speedup relative to 16 OpenMP CPU threads as a function of basis size using different MBPT methods and GPUs. The singlet closed-shell CuO$^{-}$ anion, aug-cc-pV$n$Z ($n$ = D, T, Q, and 5) basis sets, and the AMD Ryzen 9 3900X CPU with SMT enabled are used. Data is taken from Table~\ref{tab:OpenMP.OpenACC.times}.
%32 OpenMP CPU threads
%The AMD Ryzen Threadripper PRO 3975WX 32-core CPU with SMT disabled is used for OpenMP.
%NVIDIA GeForce
%The AMD Ryzen Threadripper PRO 3975WX 32-core CPU with SMT disabled and the NVIDIA GeForce RTX 4090 GPU are used for OpenMP and OpenACC, respectively.
%qs$GW$ and qsMP2 self-energy calculations

\begin{figure*}
\begin{tabular}{c}
%{\includegraphics[trim=0mm 0mm 0mm 0mm, clip, width=0.48\textwidth]{figure_openmp_openacc_speedup_CuO_gw}}
%{\includegraphics[trim=0mm 0mm 0mm 0mm, clip, width=0.48\textwidth]{figure_openmp_openacc_speedup_CuO_pt2}}
%{\includegraphics[trim=0mm 0mm 0mm 0mm, clip, width=0.48\textwidth]{figure_openmp_openacc_speedup_alkane_3_4090_gw}}
%{\includegraphics[trim=0mm 0mm 0mm 0mm, clip, width=0.48\textwidth]{figure_openmp_openacc_speedup_alkane_3_4090_pt2}}
{\includegraphics[trim=0mm 0mm 0mm 0mm, clip, width=0.48\textwidth]{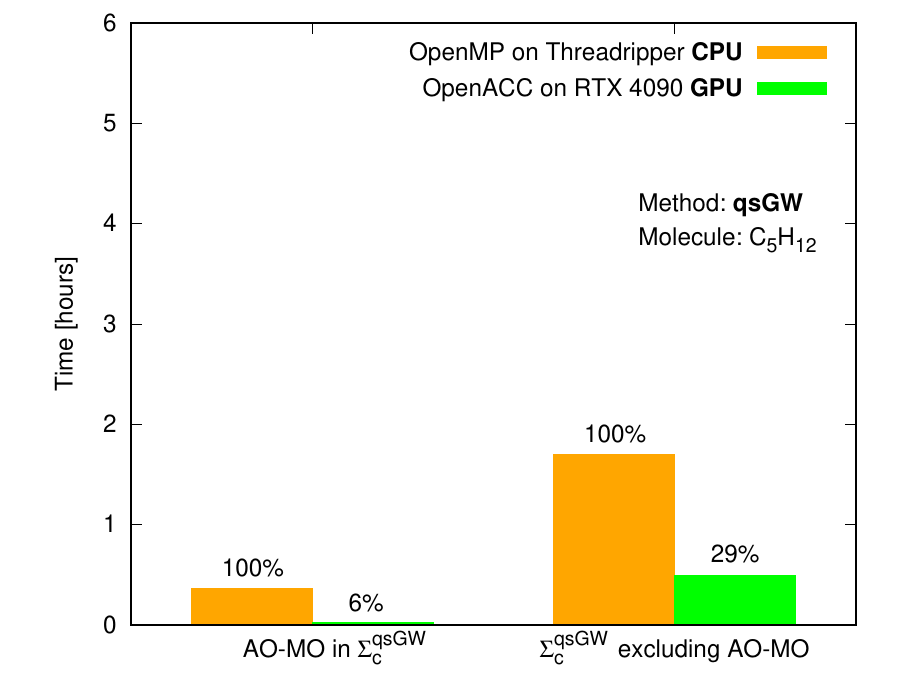}}
{\includegraphics[trim=0mm 0mm 0mm 0mm, clip, width=0.48\textwidth]{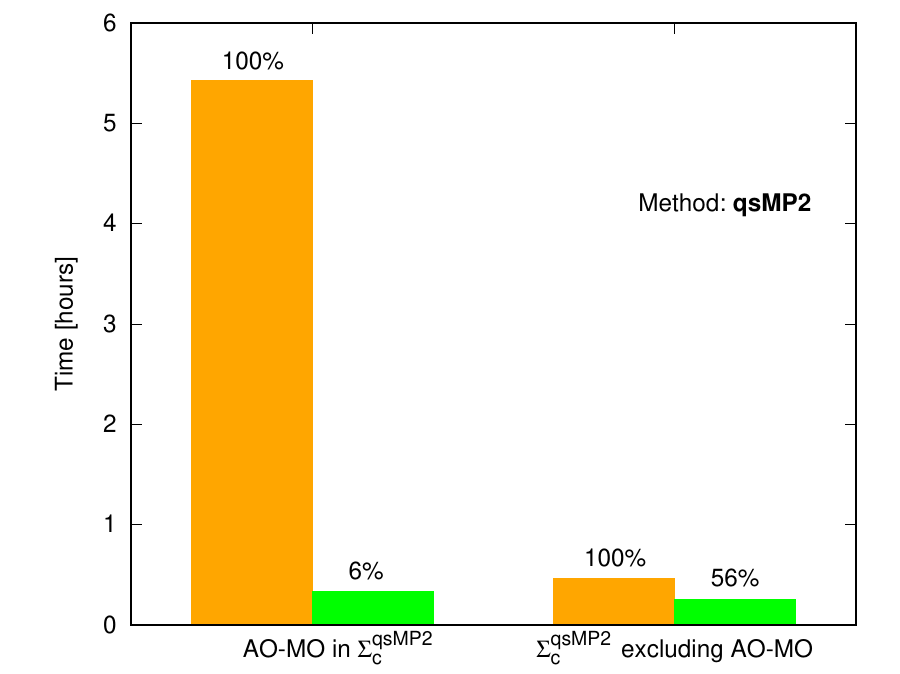}}
\end{tabular}
\caption{\textcolor{black}{(Color online) Decomposed OpenMP and OpenACC compute times of $\Sigma^{\text{qsGW}}_{\text{c}}$ and $\Sigma^{\text{qsMP2}}_{\text{c}}$ calculations (left and right, respectively) of the pentane molecule (C$_{5}$H$_{12}$). The Threadripper CPU and the RTX 4090 GPU are used for OpenMP and OpenACC, respectively. The cc-pVTZ basis set is used. AO-MO represents the atomic orbital-to-molecular orbital integral transformation. Data is taken from Tables~IV and V in Supporting Information.}}
\label{fig:bottleneck.times}
\end{figure*}

%qs$GW$ and qsMP2 self-energy
%Table~III in Supporting Information and 

\begin{figure}
\begin{tabular}{c}
%{\includegraphics[trim=0mm 0mm 0mm 0mm, clip, width=0.48\textwidth]{figure_openacc_speedup_CuO}}
%{\includegraphics[trim=0mm 0mm 0mm 0mm, clip, width=0.48\textwidth]{figure_openacc_speedup_alkanes}}
{\includegraphics[trim=0mm 0mm 0mm 0mm, clip, width=0.48\textwidth]{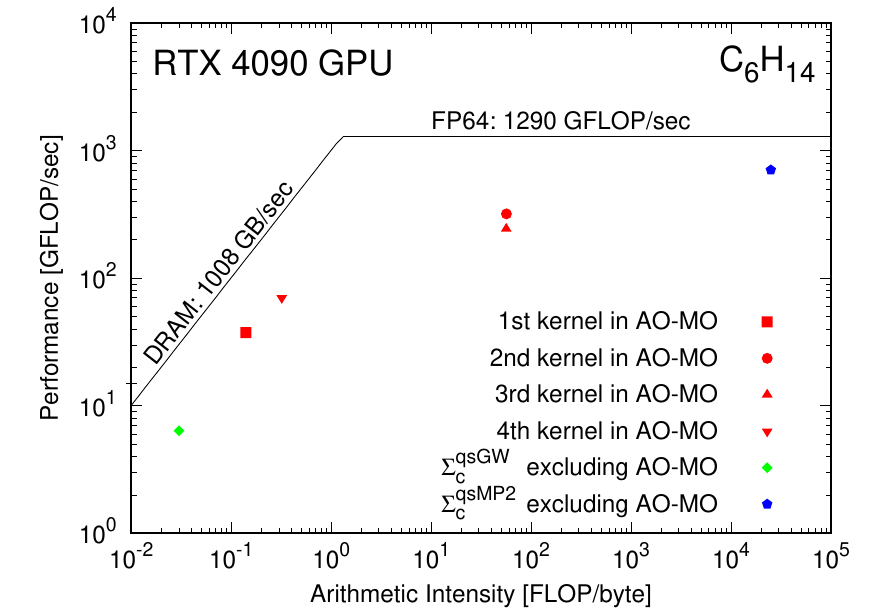}}
%{\includegraphics[trim=0mm 0mm 0mm 0mm, clip, width=0.48\textwidth]{figure_roofline_cpu}}
\end{tabular}
\caption{\textcolor{black}{(Color online) Roofline analysis of $\Sigma^{\text{qsGW}}_{\text{c}}$ and $\Sigma^{\text{qsMP2}}_{\text{c}}$ calculations of the hexane molecule (C$_{6}$H$_{14}$) on the NVIDIA GeForce RTX 4090 GPU, obtained from the NVIDIA Nsight Compute. AO-MO represents the atomic orbital-to-molecular orbital integral transformation. (G)FLOP and GB represent (giga) floating-point operations and gigabyte, respectively. DRAM and FP64 represent dynamic random access memory bandwidth and double-precision floating-point performance, respectively. Data is taken from Tables~\ref{tab:GPU.CPU.specs} and \ref{tab:roofline.GPU}.}}
\label{fig:roofline.GPU}
\end{figure}

%OpenACC speedup relative to 32 OpenMP CPU threads as a function of alkane (C$_{n}$H$_{2n+2}$) size using different MBPT methods and GPUs. The cc-pVTZ basis set is used. The AMD Ryzen Threadripper PRO 3975WX 32-core CPU with SMT disabled is used for OpenMP. Data is taken from Table~\ref{tab:OpenMP.OpenACC.times}.
%dominant bottlenecks in qs$GW$ and qsMP2 calculations

%Figure~\ref{fig:roofline.GPU} 
%of qs$GW$ and qsMP2 self-energy calculations on the RTX 4090 GPU using the hexane molecule (C$_{6}$H$_{14}$)
%We used the OpenACC version of MOLGW, the RTX 4090 GPU, the hexane molecule (C$_{6}$H$_{14}$), and the NVIDIA Nsight Compute.
%6.37~GFLOP/sec
%and assess the code performance for the computational bottlenecks
%The roofline analysis results using OpenACC calculations of the hexane molecule (C$_{6}$H$_{14}$) on the RTX 4090 GPU are shown in Fig.~\ref{fig:roofline.GPU} (the analysis results using OpenMP calculations on the Threadripper CPU are shown in Fig.~2 in Supporting Information).
%of alkane molecules (C$_{n}$H$_{2n+2}$) 
%the quad-channel memory bandwidth of 102.4~GB/s
%OpenACC
%on the GPU
%w/o

\begin{table*}
  \caption{\textcolor{black}{Dynamic random access memory (DRAM) arithmetic intensity (AI) and double-precision floating-point performance (FP64) of major computational bottlenecks in qs$GW$ and qsMP2 calculations of alkane molecules (C$_{n}$H$_{2n+2}$), obtained from the NVIDIA Nsight Compute. (G)FLOP represents (giga) floating-point operations. AO-MO represents the atomic orbital-to-molecular orbital integral transformation. $\Sigma^{\text{qsGW}}_{\text{c}}$ and $\Sigma^{\text{qsMP2}}_{\text{c}}$ represent the correlation part of qs$GW$ and qsMP2 self-energies, respectively.}}
  \label{tab:roofline.GPU}
  \begin{tabular*}{1.00\textwidth}{ @{\extracolsep{\fill}} l r r r r r r r r r r r r }
    \hline \hline
& \multicolumn{12}{c}{OpenACC (FLOP/byte for AI and GFLOP/sec for FP64)} \\
\cline{2-13}
& \multicolumn{12}{c}{NVIDIA GeForce RTX 4090 GPU} \\
\cline{2-13}
& \multicolumn{8}{c}{AO-MO} \\
\cline{2-9}
& \multicolumn{2}{c}{1st kernel} & \multicolumn{2}{c}{2nd kernel} & \multicolumn{2}{c}{3rd kernel} & \multicolumn{2}{c}{4th kernel} & \multicolumn{2}{c}{$\Sigma^{\text{qsGW}}_{\text{c}}$ excl. AO-MO} & \multicolumn{2}{c}{$\Sigma^{\text{qsMP2}}_{\text{c}}$ excl. AO-MO} \\
\cline{2-3}
\cline{4-5}
\cline{6-7}
\cline{8-9}
\cline{10-11}
\cline{12-13}
Molecule\footnotemark[1] & AI & FP64 & AI & FP64 & AI & FP64 & AI & FP64 & AI & FP64 & AI & FP64 \\
\hline
CH$_{4}$              & 1.78 & 22.45 & 35.76 & 135.3 & 36.84 & 120.2 & 36.59 & 129.8 & 0.25 & 17.36 &   4230 & 317.5 \\
C$_{2}$H$_{6}$   & 0.18 & 32.24 & 51.44 & 162.5 & 50.56 & 147.6 & 50.89 & 162.1 & 0.13 & 28.28 &   9840 & 466.6 \\
C$_{3}$H$_{8}$   & 0.13 & 33.44 & 42.04 & 210.1 & 41.34 & 188.0 & 32.91 & 144.7 & 0.07 & 12.54 & 13714 & 613.0 \\
C$_{4}$H$_{10}$ & 0.11 & 31.51 & 44.59 & 234.8 & 44.40 & 211.3 &  0.70  & 103.1 & 0.04 &   7.39 & 18102 & 604.8 \\
C$_{5}$H$_{12}$ & 0.12 & 33.11 & 49.41 & 277.4 & 49.89 & 239.2 &  0.34  &   81.1 & 0.03 &   6.56 & 21569 & 651.8 \\
C$_{6}$H$_{14}$ & 0.14 & 37.42 & 56.69 & 319.5 & 56.29 & 243.2 &  0.32  &   70.3 & 0.03 &   6.37 & 24942 & 705.3 \\
C$_{7}$H$_{16}$ & N/A\footnotemark[2] & N/A\footnotemark[2] & N/A\footnotemark[2] & N/A\footnotemark[2] & N/A\footnotemark[2] & N/A\footnotemark[2] & N/A\footnotemark[2] & N/A\footnotemark[2] & N/A\footnotemark[2] & N/A\footnotemark[2] & N/A\footnotemark[2] & N/A\footnotemark[2] \\
    \hline \hline
  \end{tabular*}
  \footnotetext[1]{See Table~\ref{tab:OpenMP.OpenACC.times} for the number of basis functions and the GPU peak memory usage}
  \footnotetext[2]{Not available because profiling calculations are prohibitively expensive. Profiling calculations take more than 10 times longer than normal calculations.}
\end{table*}

%of large molecules
%(Color online) 
%The higher utilization in each molecule and kernel is highlighted in red.

\begin{table*}
  \caption{\textcolor{black}{GPU utilization of major computational bottlenecks in qs$GW$ and qsMP2 calculations of alkane molecules (C$_{n}$H$_{2n+2}$), obtained from the NVIDIA Nsight Compute. AO-MO represents the atomic orbital-to-molecular orbital integral transformation. $\Sigma^{\text{qsGW}}_{\text{c}}$ and $\Sigma^{\text{qsMP2}}_{\text{c}}$ represent the correlation part of qs$GW$ and qsMP2 self-energies, respectively.}}
  \label{tab:utilization.GPU}
  \begin{tabular*}{1.00\textwidth}{ @{\extracolsep{\fill}} l r r r r r r r r r r r r }
    \hline \hline
& \multicolumn{12}{c}{OpenACC (\%)} \\
\cline{2-13}
& \multicolumn{12}{c}{NVIDIA GeForce RTX 4090 GPU} \\
\cline{2-13}
& \multicolumn{8}{c}{AO-MO} \\
\cline{2-9}
& \multicolumn{2}{c}{1st kernel} & \multicolumn{2}{c}{2nd kernel} & \multicolumn{2}{c}{3rd kernel} & \multicolumn{2}{c}{4th kernel} & \multicolumn{2}{c}{$\Sigma^{\text{qsGW}}_{\text{c}}$ excl. AO-MO} & \multicolumn{2}{c}{$\Sigma^{\text{qsMP2}}_{\text{c}}$ excl. AO-MO} \\
\cline{2-3}
\cline{4-5}
\cline{6-7}
\cline{8-9}
\cline{10-11}
\cline{12-13}
Molecule\footnotemark[1] & Memory & Compute & Memory & Compute & Memory & Compute & Memory & Compute & Memory & Compute & Memory & Compute \\
\hline
CH$_{4}$              & 19.2 &   7.3 & 33.0 & 81.1 & 28.7 & 69.8 & 31.4 & 76.4 & 33.1 & 1.6 & 1.4 & 38.2 \\
C$_{2}$H$_{6}$   & 29.8 & 10.5 & 30.9 & 79.4 & 38.9 & 66.2 & 62.6 & 72.9 & 75.9 & 2.5 & 2.6 & 55.9 \\
C$_{3}$H$_{8}$   & 35.6 & 10.8 & 31.0 & 79.8 & 28.9 & 69.3 & 74.4 & 53.3 & 33.6 & 1.1 & 3.8 & 73.4 \\
C$_{4}$H$_{10}$ & 37.2 &   9.9 & 30.4 & 81.2 & 64.0 & 67.6 & 57.1 & 33.0 & 19.8 & 0.7 & 3.7 & 72.4 \\
C$_{5}$H$_{12}$ & 41.1 & 10.1 & 30.5 & 81.4 & 70.7 & 66.6 & 48.7 & 22.5 & 20.0 & 0.6 & 3.9 & 77.9 \\
C$_{6}$H$_{14}$ & 42.8 & 11.1 & 30.5 & 81.4 & 93.5 & 61.9 & 42.1 & 17.9 & 20.1 & 0.6 & 4.4 & 84.3 \\
C$_{7}$H$_{16}$ & N/A\footnotemark[2] & N/A\footnotemark[2] & N/A\footnotemark[2] & N/A\footnotemark[2] & N/A\footnotemark[2] & N/A\footnotemark[2] & N/A\footnotemark[2] & N/A\footnotemark[2] & N/A\footnotemark[2] & N/A\footnotemark[2] & N/A\footnotemark[2] & N/A\footnotemark[2] \\
    \hline \hline
  \end{tabular*}
  \footnotetext[1]{See Table~\ref{tab:OpenMP.OpenACC.times} for the number of basis functions and the GPU peak memory usage}
  \footnotetext[2]{Not available because profiling calculations are prohibitively expensive. Profiling calculations take more than 10 times longer than normal calculations.}
\end{table*}

%of large molecules

\section{Results and Discussion}

\textcolor{black}{
\subsection{Performance analysis using total timings} \label{sec:total.timings}
}

%In order to assess the performance gains from GPU acceleration, we conducted an OpenMP (CPU) and OpenACC (GPU) benchmark using total timings and summarized the benchmark results in Table~\ref{tab:OpenMP.OpenACC.times}.
%\textcolor{black}{(in terms of compute time)}
%Our OpenMP (CPU) and OpenACC (GPU) benchmark results \textcolor{black}{for speedups in terms of total timings} are summarized in Table~\ref{tab:OpenMP.OpenACC.times}
%Once again, it should be noted that

\textcolor{black}{
\textcolor{black}{In order to obtain the GPU speedups over CPU, we performed an OpenMP (CPU) and OpenACC (GPU) benchmark using \emph{total timings} and summarized the benchmark results in Table~\ref{tab:OpenMP.OpenACC.times}.} We see \textcolor{black}{a possibility} that GPUs can accelerate \emph{large-scale} quasiparticle self-consistent MBPT calculations. For example, the qs$GW$/cc-pVTZ calculation of the heptane molecule (C$_{7}$H$_{16}$) on the Threadripper and Xeon Phi CPUs using only 3 qs$GW$ iterations takes \textcolor{black}{15.69} and 23.71 hours of compute time, respectively. The RTX 4090 GPU reduces the compute time to \textcolor{black}{4.98} hours (by \textcolor{black}{3.2} and 4.8 times, respectively). \textcolor{black}{As noted in Section~\ref{sec:benchmark.configurations}, about 10 times more iterations and compute times are needed for real calculations.~\cite{Bruneval12,Byun21} For example, the above calculation on the Xeon Phi CPU using 30 iterations takes about 30 days, which can exceed the walltime limit (typically, about 2 days) in HPC supercomputing centers.} In the following, we will visualize and analyze our benchmark results in Table~\ref{tab:OpenMP.OpenACC.times} to find trends.
}

%18.02 and 23.71 hours
%4.94 hours
%3.7 and 4.8 times
%using different software and hardware configurations

%Our OpenMP and OpenACC benchmark results using different software and hardware configurations are summarized in Table~\ref{tab:OpenMP.OpenACC.times}. We see that GPUs can accelerate large-scale quasiparticle self-consistent MBPT calculations. For example, the qs$GW$/aug-cc-pV5Z calculation for ZnO$^{-}$ using 16 OpenMP CPU threads takes 297 hours (about 12 days) of compute time, which can \emph{exceed} a typical walltime limit (about 2 days) in most HPC supercomputing centers. Our GPU acceleration of the calculation on RTX 3090 reduces the compute time from 297 hours to 39 hours (by 7.5 times), allowing the calculation to be finished \emph{within} the typical walltime limit. In the following, we will visualize and analyze our benchmark results in Table~\ref{tab:OpenMP.OpenACC.times} to find trends.

We begin by briefly checking the OpenMP (CPU) parallel efficiency.~\cite{Byun21} Figure~\ref{fig:OpenMP.speedup} shows the OpenMP speedup as a function of number of CPU threads using different MBPT methods. We used the propane molecule (C$_{3}$H$_{8}$) and the Threadripper CPU. We see that the qsMP2 method gives a higher parallel efficiency than the qs$GW$ one (\textcolor{black}{$\sim$0.66} and \textcolor{black}{$\sim$0.53}, respectively). This is because the implementation of $\Sigma^{\text{MP2}}_{\text{c}}$ in Eq.~(\ref{eq:Sigma.c.MP2}) takes less steps and thus is simpler than that of $\Sigma^{\text{GW}}_{\text{c}}$ in Eqs.~(\ref{eq:Casida.AB.XY}), (\ref{eq:Casida.A}), (\ref{eq:Casida.B}), (\ref{eq:Sigma.c.GW}), and (\ref{eq:w.residue}), \textcolor{black}{as shown in Fig.~\ref{fig:flowchart}. We also see that for both qs$GW$ and qsMP2 methods, the OpenMP parallel efficiency is almost independent of the number of CPU threads (the OpenMP speedup scales nearly linearly with the number of CPU threads).}

%the $\Sigma^{\text{MP2},\sigma}_{\text{c}}$ implementation
%the $\Sigma^{\text{GW},\sigma}_{\text{c}}$ one
%,\sigma
%,\sigma

%Second, the NVIDIA and Intel compilers give generally similar and constant parallel efficiencies (although the NVIDIA compilers give a slightly higher parallel efficiency than the Intel ones). This trend holds for small arrays using aug-cc-pVDZ and aug-cc-pVTZ, but for large arrays using aug-cc-pVQZ and aug-cc-pV5Z, the NVIDIA and Intel compilers give different and inconsistent parallel efficiencies possibly due to non-uniform memory access (NUMA) effects on chiplet-based CPUs, as shown in Supporting Information.

%and compilers
%We used CuO$^{-}$ and aug-cc-pVTZ. 
%($\sim$0.50 and $\sim$0.35, respectively)
%We see a couple of trends. First,

%($\sim$0.63 and $\sim$0.49, respectively)
%GPUs can make high-precision MBPT calculations of molecules more efficient than CPUs
%gives
%(\textcolor{black}{21.4x} and \textcolor{black}{85.6x}, respectively)
%(\textcolor{black}{16.5x} and \textcolor{black}{32.1x}, respectively)

Next, we compare OpenMP (CPU) and OpenACC (GPU) performances using a single alkane molecule. Figure~\ref{fig:OpenMP.OpenACC.speedup} shows OpenMP and OpenACC speedups of different MBPT methods. We used the propane molecule (C$_{3}$H$_{8}$), the Threadripper CPU, and the RTX 4090 GPU. We see a couple of trends. First, for both qs$GW$ and qsMP2 methods, OpenACC \textcolor{black}{achieves} higher speedups than 32 OpenMP threads (by \textcolor{black}{1.9x} and \textcolor{black}{4.0x}, respectively), which shows that GPUs can \textcolor{black}{perform} high-precision MBPT calculations of molecules \textcolor{black}{faster} than CPUs. Second, for both OpenMP and OpenACC, the qsMP2 method gets higher speedups than the qs$GW$ one (by \textcolor{black}{1.3x} and \textcolor{black}{2.7x}, respectively), \textcolor{black}{which shows that the performance boost from parallelization depends on the kind of MBPT methods.}

%because it is easier to implement and parallelize $\Sigma^{\text{MP2}}_{\text{c}}$ than $\Sigma^{\text{GW}}_{\text{c}}$, as explained above.
%which will be discussed in detail in Section

%2.2x and 5.3x
%34.8x and 106.5x
%15.8x and 20.2x

%,\sigma
%,\sigma

%a fixed basis size
%We used CuO$^{-}$ and RTX 3090. We used aug-cc-pVQZ, because aug-cc-pV5Z calculations using a single CPU thread are not computationally feasible, as shown in Table~\ref{tab:OpenMP.OpenACC.times}.
%(2.6x and 6.2x, respectively)
%(9.1x and 56.3x, respectively)
%(6.4x and 16.7x, respectively)
%These trends hold for ZnO$^{-}$ as well, as shown in Supporting Information.

%Last,
%and the Threadripper CPU
%as discussed above.
%2.7x and 3.7x

%For example, for butane and heptane molecules (C$_{4}$H$_{10}$ and C$_{7}$H$_{16}$, respectively), qs$GW$ calculations on the RTX 4090 use 5.0 and 21.6~GB of GPU memory, respectively, and get OpenACC speedups of 2.7x and 3.7x, respectively, as shown in Table~\ref{tab:OpenMP.OpenACC.times}.

%need
%0.7x
%give
%1.4 and 2.2 times
%the qsMP2 calculation of C$_{7}$H$_{16}$ with 21.4~GB of GPU memory footprint on the RTX 4090 gives the highest OpenACC speedup of 13.6x.
%13.6x
%We used the cc-pVTZ basis set. 
%because they do not have a sufficient number of threads to saturate CUDA cores
%\textcolor{black}{due to an insufficient}

\textcolor{black}{Finally,} we analyze the effect of the molecule size on the OpenACC (GPU) performance. Figure~\ref{fig:OpenACC.speedup} shows the OpenACC \textcolor{black}{GPU} speedup with respect to OpenMP CPU threads as a function of alkane (C$_{n}$H$_{2n+2}$) size using different MBPT methods, GPUs, \textcolor{black}{and CPUs}. We see a few trends. First, the OpenACC speedup depends on the kind of MBPT methods: for all alkane sizes and GPUs, the qsMP2 method gets higher OpenACC speedups than the qs$GW$ one. \textcolor{black}{This trend will be discussed in detail in Section~\ref{sec:partial.timings}.}

Second, the OpenACC speedup increases with the alkane size (the basis size or the GPU memory footprint):~\cite{Isborn11,Pham23} for all MBPT methods and GPUs, larger alkane molecules \textcolor{black}{get} higher OpenACC speedups, because large arrays utilize more CUDA cores \textcolor{black}{than small arrays}. For example, qs$GW$ calculations of butane and heptane molecules (C$_{4}$H$_{10}$ and C$_{7}$H$_{16}$, respectively) on the RTX 4090 GPU use 5.0 and 21.6~GB of GPU memory, respectively, and get OpenACC speedups of \textcolor{black}{2.2x} and \textcolor{black}{3.2x}, respectively, \textcolor{black}{compared to those on the Threadripper CPU}, as shown in Table~\ref{tab:OpenMP.OpenACC.times}. Also, \emph{small-scale} MBPT calculations using small basis sets might not \textcolor{black}{benefit from} GPU acceleration, because they do not have a sufficient number of threads to saturate CUDA cores. \textcolor{black}{For example,} the qs$GW$ calculation of the methane molecule (CH$_{4}$), \textcolor{black}{the smallest alkane}, on the RTX 4090 GPU uses \textcolor{black}{only} 86 basis functions and gets an OpenACC speedup of \textcolor{black}{0.5x relative to that on the Threadripper CPU (the calculation runs faster on the CPU than on the GPU)}, as shown in Table~\ref{tab:OpenMP.OpenACC.times}. \textcolor{black}{This trend will be discussed in detail in Section~\ref{sec:roofline}.}

\textcolor{black}{
We analyzed the asymptotic time complexity of the OpenACC speedup with respect to alkane (C$_{n}$H$_{2n+2}$) size using a fitting function:
\begin{equation}
f(n) = \alpha \times n^{\beta}, \label{eq:fit}
\end{equation}
where $\alpha$ and $\beta$ are fitting constants, and summarized the analysis results in Table~VI in Supporting Information. For example, we obtained $f(n) = 1.30 \times n^{1.05}$ for the qsMP2 method on the RTX 4090 GPU and the Threadripper CPU, as shown in Fig.~\ref{fig:OpenACC.speedup}. The scaling factor $\beta$ ranges from 0.56 (the qs$GW$ method on the RTX 4090 GPU and the Xeon Phi CPU) to 1.21 (the qs$GW$ method on the RTX 3090 GPU and the Threadripper CPU). 
}

%presented
%K^{\sigma}_{\mu\nu} = \sum_{\lambda\tau} D^{\sigma}_{\lambda\tau} (\mu\lambda|\tau\nu). \label{eq:K}
%Figure~\ref{fig:OpenACC.speedup} shows
%qsMP2 method on the RTX 4090 GPU
%estimated
%by 2.0 times
%by \textcolor{black}{1.9} and \textcolor{black}{2.4} times

%How does your code scale with respect to the n?

\textcolor{black}{
Third, the OpenACC speedup depends on the kind of CPUs: for all MBPT methods and alkane sizes, the Xeon Phi CPU underperforms the Threadripper CPU, taking longer OpenMP compute times and thus giving higher OpenACC speedups. For example, for the qsMP2 calculation of C$_{4}$H$_{10}$, the Xeon Phi CPU gives a 2.0x higher OpenACC speedup than the Threadripper CPU, as shown in Table~\ref{tab:OpenMP.OpenACC.times}.
}

Last, the OpenACC speedup increases with the GPU generation: for all MBPT methods and alkane sizes, the RTX 4090 \textcolor{black}{GPU takes shorter OpenACC compute times and thus} gives higher OpenACC speedups than the RTX 3090 \textcolor{black}{GPU} (see Table~\ref{tab:GPU.CPU.specs} for their specifications). For example, for qs$GW$ and qsMP2 calculations of C$_{4}$H$_{10}$, the RTX 4090 \textcolor{black}{GPU} gives higher OpenACC speedups than the RTX 3090 \textcolor{black}{GPU} by \textcolor{black}{1.9} and \textcolor{black}{2.4} times, respectively, \textcolor{black}{as shown in Table~\ref{tab:OpenMP.OpenACC.times}}, which are close to the ratio of the RTX 4090 FP64 performance to the RTX 3090 one (1290~GFLOP/s $\div$ 556~GFLOP/s $\approx$ 2.3), as shown in Table~\ref{tab:GPU.CPU.specs}.

Overall, the qsMP2 calculation of C$_{7}$H$_{16}$ on the RTX 4090 GPU using 21.4~GB of GPU memory \textcolor{black}{achieves} the highest OpenACC speedup of \textcolor{black}{9.7x and 14.7x with respect to that on the Threadripper and Xeon Phi CPUs, respectively, as shown in Table~\ref{tab:OpenMP.OpenACC.times}}. Based on the trends above, we expect that larger-memory-footprint MBPT calculations on high-end GPUs, such as the NVIDIA H100 with a high FP64 performance and 80~GB of an extension of the second generation of high bandwidth memory (HBM2e), would \textcolor{black}{achieve} higher OpenACC speedups than \textcolor{black}{9.7x and 14.7x}.

\textcolor{black}{
\subsection{Performance analysis using partial timings} \label{sec:partial.timings}
}

%the qsMP2 method benefits more from GPU acceleration than the qs$GW$ method
%In Listing, the AO-MO bottleneck corresponds to a line 1, and the $\Sigma^{\text{qsMP2}}_{\text{c}}$ without AO-MO bottleneck corresponds to lines 1--2, as explained in Section.
%We used the cc-pVTZ basis set.
%three major bottlenecks (AO-MO, $\Sigma^{\text{qsGW}}_{\text{c}}$ without AO-MO, and $\Sigma^{\text{qsMP2}}_{\text{c}}$ without AO-MO, as shown in Table~\ref{tab:OpenACC.directives.clauses})
%we analyzed partial timings in our OpenMP (CPU) and OpenACC (GPU) benchmark results and summarized the analysis results in Tables~VI and V in Supporting Information.
%$\Sigma^{\text{qsGW}}_{\text{c}}$
%for OpenMP (CPU)
%4.7 times
%11.8 times

\textcolor{black}{
In order to find why the qsMP2 method gets higher OpenACC speedups than the qs$GW$ method, as seen in Section~\ref{sec:total.timings}, we performed an OpenMP (CPU) and OpenACC (GPU) benchmark using \emph{partial timings} and summarized the benchmark results in Tables~IV and V in Supporting Information. As explained in Section~\ref{sec:OpenACC.implementation}, the $\Sigma^{\text{qsMP2}}_{\text{c}}$ calculation is composed of two computational bottlenecks: the AO-MO integral transformation (AO-MO in the following) and the AO-MO-excluded $\Sigma^{\text{qsMP2}}_{\text{c}}$ (see Table~\ref{tab:OpenACC.directives.clauses} and Listing~\ref{list:source.code}), which is the case for the $\Sigma^{\text{qsGW}}_{\text{c}}$ calculation as well. Figure~\ref{fig:bottleneck.times} shows decomposed OpenMP (CPU) and OpenACC (GPU) compute times of $\Sigma^{\text{qsGW}}_{\text{c}}$ and $\Sigma^{\text{qsMP2}}_{\text{c}}$ calculations of the pentane molecule (C$_{5}$H$_{12}$) on the Threadripper CPU and the RTX 4090 GPU. We see that in the case of OpenMP (CPU) calculations, the AO-MO bottleneck does not dominate the $\Sigma^{\text{qsGW}}_{\text{c}}$ calculation, but does the $\Sigma^{\text{qsMP2}}_{\text{c}}$ calculation. For example, the AO-MO bottleneck takes 4.7 times \emph{less} OpenMP compute time than the AO-MO-excluded $\Sigma^{\text{qsGW}}_{\text{c}}$ bottleneck in the $\Sigma^{\text{qsGW}}_{\text{c}}$ calculation, but 11.8 times \emph{more} OpenMP compute time than the AO-MO-excluded $\Sigma^{\text{qsMP2}}_{\text{c}}$ bottleneck in the $\Sigma^{\text{qsMP2}}_{\text{c}}$ calculation (see Table~IV in Supporting Information). We also see that the AO-MO bottleneck benefits the most from GPU acceleration among three major computational bottlenecks. For example, OpenACC (GPU) reduces OpenMP (CPU) compute times of AO-MO, AO-MO-excluded $\Sigma^{\text{qsGW}}_{\text{c}}$, and AO-MO-excluded $\Sigma^{\text{qsMP2}}_{\text{c}}$ bottlenecks by 94\%, 71\%, and 46\%. As a result, GPU acceleration has a stronger effect on the qsMP2 method than the qs$GW$ method (GPU speedups of 8.9x and 3.2x, respectively, as shown in Table~\ref{tab:OpenMP.OpenACC.times}).
}	

%The AMD Ryzen Threadripper PRO 3975WX 32-core CPU with SMT disabled and the NVIDIA GeForce RTX 4090 GPU are used for OpenMP and OpenACC, respectively. The cc-pVTZ basis set is used.
%Our OpenMP (CPU) and OpenACC (GPU) benchmark results using total timings are summarized in Table~\ref{tab:OpenMP.OpenACC.times}.

\textcolor{black}{
\subsection{Roofline performance analysis} \label{sec:roofline}
}

%the MOLGW code
%dominant
%the AO-MO integral transformation consisting of 4 kernels, $\Sigma^{\text{qsGW}}_{\text{c}}$ without the AO-MO transformation, and $\Sigma^{\text{qsMP2}}_{\text{c}}$ without the AO-MO transformation
%summarizes arithmetic intensity (AI) and FP64 performance of the 6 kernels in OpenACC calculations of alkane molecules (C$_{n}$H$_{2n+2}$) on the RTX 4090 GPU

\textcolor{black}{
In order to find what is limiting the performance of our OpenACC implementation, we conducted the roofline performance analysis~\cite{Williams09} of three major computational bottlenecks (a total of 6 kernels) -- AO-MO (consisting of 4 kernels), AO-MO-excluded $\Sigma^{\text{qsGW}}_{\text{c}}$, and AO-MO-excluded $\Sigma^{\text{qsMP2}}_{\text{c}}$ (see Table~\ref{tab:OpenACC.directives.clauses}) -- in OpenACC calculations on the RTX 4090 GPU, and summarized the analysis results in Table~\ref{tab:roofline.GPU}. Figure~\ref{fig:roofline.GPU} depicts the roofline analysis results for the hexane molecule (C$_{6}$H$_{14}$). We see that the AO-MO bottleneck, which is common in qs$GW$ and qsMP2 calculations, has both compute- and memory-bound kernels. For example, the first and fourth kernels in the AO-MO bottleneck are under the peak bandwidth ceiling (diagnoal line) and thus are memory-bound, while the second and third kernels are under the peak performance ceiling (horizontal line) and thus are compute-bound. We also see that the AO-MO-excluded $\Sigma^{\text{qsGW}}_{\text{c}}$ bottleneck is highly memory-bound (arithmetic intensity of 0.03~FLOP/byte), whereas the AO-MO-excluded $\Sigma^{\text{qsMP2}}_{\text{c}}$ bottleneck is highly compute-bound (FP64 performance of 705.3~GFLOP/sec) (see Table~\ref{tab:roofline.GPU}). This shows that the qs$GW$ method is more memory-bound and less compute-bound than the qsMP2 method. We summarized the roofline analysis results for OpenMP calculations on the Threadripper CPU in Table~VII in Supporting Information, and depicted the results for C$_{6}$H$_{14}$ in Fig.~2 in Supporting Information.
}

%the code optimization level
%of alkane molecules (C$_{n}$H$_{2n+2}$)
%depends on the kernel
%are summarized Table~I in Supporting Information and show that unlike the GPU utilization, the CPU utilization is almost independent of the molecule size.
%across all molecule sizes

\textcolor{black}{
In order to understand the optimization level of our OpenACC implementation, we analyzed the GPU utilization of three major computational bottlenecks (a total of 6 kernels) in OpenACC calculations on the RTX 4090 GPU with the FP64 performance of 1290~GFLOP/s and the memory bandwidth of 1008~GB/s (see Table~\ref{tab:GPU.CPU.specs}), and summarized the analysis results in Table~\ref{tab:utilization.GPU}. We see that the GPU utilization depends on the molecule size. For example, the GPU compute utilization of the AO-MO-excluded $\Sigma^{\text{qsMP2}}_{\text{c}}$ bottleneck increases with the alkane size from 38.2\% to 84.3\% (CH$_{4}$ and C$_{7}$H$_{16}$, respectively), while that of the second kernel in the AO-MO bottleneck remains nearly constant at $\sim$80\% across all alkane sizes. Also, the GPU memory utilization of the first kernel in the AO-MO bottleneck increases with the molecule size from 19.2\% to 42.8\% (CH$_{4}$ and C$_{7}$H$_{16}$, respectively), whereas that of the second kernel remains almost unchanged at $\sim$30\% across all alkane sizes. We also see that the GPU utilization depends on the kind of kernels. For example, in the case of the hexane molecule (C$_{6}$H$_{14}$), the GPU compute utilization ranges from 0.6\% to 84.3\% (AO-MO-excluded $\Sigma^{\text{qsGW}}_{\text{c}}$ and $\Sigma^{\text{qsMP2}}_{\text{c}}$ kernels, respectively), and the GPU memory utilization ranges from 4.4\% to 93.5\% (the AO-MO-excluded $\Sigma^{\text{qsMP2}}_{\text{c}}$ kernel and the third kernel in the AO-MO bottleneck, respectively). We summarized the analysis results for the CPU utilization of three major computational bottlenecks in OpenMP calculations on the Threadripper CPU with the measured FP64 performance of 268~GFLOP/s and the measured quad-channel memory bandwidth of 61.2~GB/s (see Table~\ref{tab:GPU.CPU.specs}) in Table~VIII in Supporting Information. 
}

\textcolor{black}{
\subsection{Discussion} \label{sec:discussion}
}

%\textcolor{black}{
We have a few points to discuss. First, OpenMP and OpenACC implementations in MOLGW ``added parallelism into existing source code without significantly modifying it,''~\cite{GuideOpenMP} enabling us to maintain a \emph{single} version of source code for both CPUs and GPUs. The clean and maintainable MOLGW source code makes it possible for domain scientists, such as electronic structure method developers, with no or little parallel programming background to do ``more science and less programming.''~\cite{OpenMP}
%}

%We have a few points to discuss. First, as the creator of the Python programming language said ``Maintainable code is more important than clever code,''~\cite{Dropbox} maintainability is as important as performance in software development including the development of HPC scientific applications. OpenMP and OpenACC implementations in MOLGW ``added parallelism into existing source code without significantly modifying it,''~\cite{GuideOpenMP} enabling us to maintain a single version of source code for both CPUs and GPUs. The clean and maintainable MOLGW source code makes it possible for domain scientists, such as electronic structure method developers, with no or little parallel programming background to do ``more science and less programming.''~\cite{OpenMP}

%in the near future

Second, our current OpenACC version of MOLGW runs only on a single GPU. We will enable MOLGW to run on a multi-GPU node using the hybrid OpenMP/OpenACC parallelization in the future. The multi-GPU implementation can enhance scalability of MOLGW,~\cite{Barca21} but will reduce readability, understandability, and backward compatibility of source code, because it requires to change the serial CPU code.

%Third, we did not add OpenACC to the MPI-parallelized RI approximation in MOLGW. We are planning to GPU accelerate the RI approximation with the hybrid MPI/OpenACC parallelization.

%Fourth, we inserted OpenACC into starting-point-\emph{dependent} MBPT methods, such as $G_{0}W_{0}$, as well. We will report our benchmark results for them elsewhere.

%Last, we ported $GW$ and MP2 methods for electronic excitations to the GPU. It is possible to port Bethe--Salpeter equation (BSE) and time-dependent density-functional theory (TDDFT) methods for optical excitations to the GPU as well.

%with $N$ being the system size
%However, it will become a major computational bottleneck when very large systems are used.

\textcolor{black}{
Third, the matrix diagonalization in MOLGW is not ported to the GPU, as discussed in Section~\ref{sec:difference.OpenMP.OpenACC}. Although the diagonalization of the Casida matrix scales as $O(N^6)$, it is a minor computational bottleneck in this work due to a small prefactor, as discussed in Section~\ref{sec:bottlenecks}. However, it will become a major computational bottleneck in $G_{0}W_{0}$ and qs$GW$ calculations of very large systems. Hence, we will port the matrix diagonalization in MOLGW to the GPU in the future possibly using a GPU-accelerated library for linear algebra.
}

%For example, $G_{0}W_{0}$ and MP2@HF methods need only diagonal elements of Eqs.~(\ref{eq:Sigma.c.GW}) and (\ref{eq:Sigma.c.MP2}) ($m = n$) and no SCF iterations for Eqs.~(\ref{eq:Sigma.c.qsGW.real}) and (\ref{eq:Sigma.c.qsMP2.real}), and thus should attain lower GPU speedups than qs$GW$ and qsMP2 ones.
%@HF
%as discussed above
%very

Last, not all MBPT methods and molecular systems would benefit from GPU acceleration. \textcolor{black}{For example, computationally cheap one-shot $G_{0}W_{0}$ and MP2@HF methods should attain lower GPU speedups than computationally expensive qs$GW$ and qsMP2 ones, as discussed in Section~\ref{sec:costs}.} Also, MBPT calculations of small systems could run faster on the CPU than on the GPU, \textcolor{black}{as seen in Section~\ref{sec:partial.timings}}, and those of large systems might require more memory than available on the GPU (typically, the CPU has more memory than the GPU, \textcolor{black}{as shown in Table~\ref{tab:GPU.CPU.specs}}). Overall, the GPU version of a MBPT code cannot fully replace the CPU version, because CPUs are still needed for certain kinds of MBPT methods and molecular systems.

%enabling them to benefit from the GPU porting. 
%GPU acceleration is not suitable for all MBPT methods and system sizes
%MBPT calculations for very large systems
%13.6 times

\section{Summary and Conclusions}

In summary, we have ported MOLGW to the GPU using OpenACC without sacrificing accuracy, and evaluated the performance of GPU-accelerated MOLGW using different starting-point-independent MBPT methods, system sizes, and GPUs. We found that the GPU-accelerated version of MOLGW can run faster than the CPU version using 32 OpenMP threads by up to \textcolor{black}{9.7} times, and the speedup depends on the kind of MBPT methods and increases with the GPU generation and the system size (the basis size or the GPU memory footprint). \textcolor{black}{We identified both compute- and memory-bound kernels in GPU-accelerated MOLGW using the roofline performance analysis.} Our choice of OpenACC over CUDA allows us to maintain a single version of the MOLGW source code for both CPUs and GPUs, which significantly reduces programming and maintenance efforts and enhances code and performance portability. Our GPU acceleration of quasiparticle self-consistent MBPT methods in MOLGW paves the way for the application of \textit{ab initio} MBPT calculations without empirical parameters, such as starting points, to complex real systems.

%basis sets
%low-end GPUs
%on desktop GPUs

%\caption{(Color online) OpenMP and OpenACC speedups of qs$GW$ and qsMP2 methods (left and right, respectively). The propane molecule (C$_{3}$H$_{8}$) and the cc-pVTZ basis set are used. The AMD Ryzen Threadripper PRO 3975WX 32-core CPU with SMT disabled and the NVIDIA GeForce RTX 4090 GPU are used for OpenMP and OpenACC, respectively. GB represents the gigabyte. Data is taken from Table~\ref{tab:OpenMP.OpenACC.times}.

%the MOLGW code

\section{Authorship Contribution Statement}

Young-Moo Byun conceived and designed the project, implemented and validated the computer code, performed the computations, collected the data, analyzed and interpreted the results, and wrote the draft manuscript. Jejoong Yoo supervised the project, provided the computing resources, and reviewed and revised the manuscript. All authors reviewed the results and approved the final version of the manuscript.

\section{Data Availability Statement}

Basis sets and the MOLGW source code with our OpenACC implementation for a single GPU are publicly available via GitHub at \url{https://github.com/ymbyun/molgw-1.F-openmp-openacc-single-gpu}.

%\section{Supporting Information}

%Additional Supporting Information may be found in the online version of this article.

% If you have acknowledgments, this puts in the proper section head.
\begin{acknowledgments}
% Put your acknowledgments here.
This work was supported by the National Research Foundation of Korea (NRF) grant funded by the Korea government (MSIT) (No. 2020R1A2C1101424). This work was supported by Institute for Information \& Communications Technology Promotion (IITP) grant funded by the Korea government (MSIT) (No. 2021-0-02068, Artificial Intelligence Innovation Hub). This work was supported by the National Supercomputing Center with supercomputing resources including technical support (KSC-2021-CRE-0212).
\end{acknowledgments}

%This work was supported by the U.S. Department of Energy Grant No. DE-SC0017824. This work was supported by the National Research Foundation of Korea (NRF) grant funded by the Korea government (MSIT) (No. 2020R1A2C1101424). The computation for this work was performed on the high performance computing infrastructure provided by Research Computing Support Services at the University of Missouri-Columbia. The authors gratefu​lly acknowledge the Advanced Cyberinfrastructure for Education and Research (ACER) group at the University of Illinois at Chicago for providing computational resources and services needed to deliver research results delivered within this paper. URL: https://acer.uic.edu This work was supported by the National Supercomputing Center with supercomputing resources including technical support (KSC-2020-CRE-0080).

% Create the reference section using BibTeX:
\bibliography{GPU}

%merlin.mbs apsrev4-1.bst 2010-07-25 4.21a (PWD, AO, DPC) hacked
%Control: key (0)
%Control: author (8) initials jnrlst
%Control: editor formatted (1) identically to author
%Control: production of article title (-1) disabled
%Control: page (0) single
%Control: year (1) truncated
%Control: production of eprint (0) enabled
\begin{thebibliography}{56}%
\makeatletter
\providecommand \@ifxundefined [1]{%
 \@ifx{#1\undefined}
}%
\providecommand \@ifnum [1]{%
 \ifnum #1\expandafter \@firstoftwo
 \else \expandafter \@secondoftwo
 \fi
}%
\providecommand \@ifx [1]{%
 \ifx #1\expandafter \@firstoftwo
 \else \expandafter \@secondoftwo
 \fi
}%
\providecommand \natexlab [1]{#1}%
\providecommand \enquote  [1]{``#1''}%
\providecommand \bibnamefont  [1]{#1}%
\providecommand \bibfnamefont [1]{#1}%
\providecommand \citenamefont [1]{#1}%
\providecommand \href@noop [0]{\@secondoftwo}%
\providecommand \href [0]{\begingroup \@sanitize@url \@href}%
\providecommand \@href[1]{\@@startlink{#1}\@@href}%
\providecommand \@@href[1]{\endgroup#1\@@endlink}%
\providecommand \@sanitize@url [0]{\catcode `\\12\catcode `\$12\catcode
  `\&12\catcode `\#12\catcode `\^12\catcode `\_12\catcode `\%12\relax}%
\providecommand \@@startlink[1]{}%
\providecommand \@@endlink[0]{}%
\providecommand \url  [0]{\begingroup\@sanitize@url \@url }%
\providecommand \@url [1]{\endgroup\@href {#1}{\urlprefix }}%
\providecommand \urlprefix  [0]{URL }%
\providecommand \Eprint [0]{\href }%
\providecommand \doibase [0]{http://dx.doi.org/}%
\providecommand \selectlanguage [0]{\@gobble}%
\providecommand \bibinfo  [0]{\@secondoftwo}%
\providecommand \bibfield  [0]{\@secondoftwo}%
\providecommand \translation [1]{[#1]}%
\providecommand \BibitemOpen [0]{}%
\providecommand \bibitemStop [0]{}%
\providecommand \bibitemNoStop [0]{.\EOS\space}%
\providecommand \EOS [0]{\spacefactor3000\relax}%
\providecommand \BibitemShut  [1]{\csname bibitem#1\endcsname}%
\let\auto@bib@innerbib\@empty
%</preamble>
\bibitem [{\citenamefont {Onida}\ \emph {et~al.}(2002)\citenamefont {Onida},
  \citenamefont {Reining},\ and\ \citenamefont {Rubio}}]{Onida02}%
  \BibitemOpen
  \bibfield  {author} {\bibinfo {author} {\bibfnamefont {G.}~\bibnamefont
  {Onida}}, \bibinfo {author} {\bibfnamefont {L.}~\bibnamefont {Reining}}, \
  and\ \bibinfo {author} {\bibfnamefont {A.}~\bibnamefont {Rubio}},\ }\href
  {\doibase 10.1103/RevModPhys.74.601} {\bibfield  {journal} {\bibinfo
  {journal} {Rev. Mod. Phys.}\ }\textbf {\bibinfo {volume} {74}},\ \bibinfo
  {pages} {601} (\bibinfo {year} {2002})}\BibitemShut {NoStop}%
\bibitem [{\citenamefont {Shishkin}\ and\ \citenamefont
  {Kresse}(2007)}]{Shishkin07a}%
  \BibitemOpen
  \bibfield  {author} {\bibinfo {author} {\bibfnamefont {M.}~\bibnamefont
  {Shishkin}}\ and\ \bibinfo {author} {\bibfnamefont {G.}~\bibnamefont
  {Kresse}},\ }\href {\doibase 10.1103/PhysRevB.75.235102} {\bibfield
  {journal} {\bibinfo  {journal} {Phys. Rev. B}\ }\textbf {\bibinfo {volume}
  {75}},\ \bibinfo {pages} {235102} (\bibinfo {year} {2007})}\BibitemShut
  {NoStop}%
\bibitem [{\citenamefont {Govoni}\ and\ \citenamefont
  {Galli}(2015)}]{Govoni15}%
  \BibitemOpen
  \bibfield  {author} {\bibinfo {author} {\bibfnamefont {M.}~\bibnamefont
  {Govoni}}\ and\ \bibinfo {author} {\bibfnamefont {G.}~\bibnamefont {Galli}},\
  }\href {\doibase 10.1021/ct500958p} {\bibfield  {journal} {\bibinfo
  {journal} {Journal of Chemical Theory and Computation}\ }\textbf {\bibinfo
  {volume} {11}},\ \bibinfo {pages} {2680} (\bibinfo {year} {2015})},\ \bibinfo
  {note} {pMID: 26575564},\ \Eprint
  {http://arxiv.org/abs/https://doi.org/10.1021/ct500958p}
  {https://doi.org/10.1021/ct500958p} \BibitemShut {NoStop}%
\bibitem [{\citenamefont {van Setten}\ \emph {et~al.}(2015)\citenamefont {van
  Setten}, \citenamefont {Caruso}, \citenamefont {Sharifzadeh}, \citenamefont
  {Ren}, \citenamefont {Scheffler}, \citenamefont {Liu}, \citenamefont
  {Lischner}, \citenamefont {Lin}, \citenamefont {Deslippe}, \citenamefont
  {Louie}, \citenamefont {Yang}, \citenamefont {Weigend}, \citenamefont
  {Neaton}, \citenamefont {Evers},\ and\ \citenamefont {Rinke}}]{vanSetten15}%
  \BibitemOpen
  \bibfield  {author} {\bibinfo {author} {\bibfnamefont {M.~J.}\ \bibnamefont
  {van Setten}}, \bibinfo {author} {\bibfnamefont {F.}~\bibnamefont {Caruso}},
  \bibinfo {author} {\bibfnamefont {S.}~\bibnamefont {Sharifzadeh}}, \bibinfo
  {author} {\bibfnamefont {X.}~\bibnamefont {Ren}}, \bibinfo {author}
  {\bibfnamefont {M.}~\bibnamefont {Scheffler}}, \bibinfo {author}
  {\bibfnamefont {F.}~\bibnamefont {Liu}}, \bibinfo {author} {\bibfnamefont
  {J.}~\bibnamefont {Lischner}}, \bibinfo {author} {\bibfnamefont
  {L.}~\bibnamefont {Lin}}, \bibinfo {author} {\bibfnamefont {J.~R.}\
  \bibnamefont {Deslippe}}, \bibinfo {author} {\bibfnamefont {S.~G.}\
  \bibnamefont {Louie}}, \bibinfo {author} {\bibfnamefont {C.}~\bibnamefont
  {Yang}}, \bibinfo {author} {\bibfnamefont {F.}~\bibnamefont {Weigend}},
  \bibinfo {author} {\bibfnamefont {J.~B.}\ \bibnamefont {Neaton}}, \bibinfo
  {author} {\bibfnamefont {F.}~\bibnamefont {Evers}}, \ and\ \bibinfo {author}
  {\bibfnamefont {P.}~\bibnamefont {Rinke}},\ }\href {\doibase
  10.1021/acs.jctc.5b00453} {\bibfield  {journal} {\bibinfo  {journal} {Journal
  of Chemical Theory and Computation}\ }\textbf {\bibinfo {volume} {11}},\
  \bibinfo {pages} {5665} (\bibinfo {year} {2015})},\ \bibinfo {note} {pMID:
  26642984},\ \Eprint
  {http://arxiv.org/abs/https://doi.org/10.1021/acs.jctc.5b00453}
  {https://doi.org/10.1021/acs.jctc.5b00453} \BibitemShut {NoStop}%
\bibitem [{\citenamefont {Byun}\ and\ \citenamefont
  {\"{O}\u{g}\"{u}t}(2019)}]{Byun19}%
  \BibitemOpen
  \bibfield  {author} {\bibinfo {author} {\bibfnamefont {Y.-M.}\ \bibnamefont
  {Byun}}\ and\ \bibinfo {author} {\bibfnamefont {S.}~\bibnamefont
  {\"{O}\u{g}\"{u}t}},\ }\href {\doibase 10.1063/1.5118671} {\bibfield
  {journal} {\bibinfo  {journal} {The Journal of Chemical Physics}\ }\textbf
  {\bibinfo {volume} {151}},\ \bibinfo {pages} {134305} (\bibinfo {year}
  {2019})},\ \Eprint {http://arxiv.org/abs/https://doi.org/10.1063/1.5118671}
  {https://doi.org/10.1063/1.5118671} \BibitemShut {NoStop}%
\bibitem [{\citenamefont {Bruneval}\ and\ \citenamefont
  {Marques}(2013)}]{Bruneval13}%
  \BibitemOpen
  \bibfield  {author} {\bibinfo {author} {\bibfnamefont {F.}~\bibnamefont
  {Bruneval}}\ and\ \bibinfo {author} {\bibfnamefont {M.~A.~L.}\ \bibnamefont
  {Marques}},\ }\href {\doibase 10.1021/ct300835h} {\bibfield  {journal}
  {\bibinfo  {journal} {Journal of Chemical Theory and Computation}\ }\textbf
  {\bibinfo {volume} {9}},\ \bibinfo {pages} {324} (\bibinfo {year} {2013})},\
  \bibinfo {note} {pMID: 26589035},\ \Eprint
  {http://arxiv.org/abs/https://doi.org/10.1021/ct300835h}
  {https://doi.org/10.1021/ct300835h} \BibitemShut {NoStop}%
\bibitem [{\citenamefont {Byun}\ \emph {et~al.}()\citenamefont {Byun},
  \citenamefont {Yoo},\ and\ \citenamefont {\"{O}\u{g}\"{u}t}}]{Byun21}%
  \BibitemOpen
  \bibfield  {author} {\bibinfo {author} {\bibfnamefont {Y.-M.}\ \bibnamefont
  {Byun}}, \bibinfo {author} {\bibfnamefont {J.}~\bibnamefont {Yoo}}, \ and\
  \bibinfo {author} {\bibfnamefont {S.}~\bibnamefont {\"{O}\u{g}\"{u}t}},\
  }\href@noop {} {\enquote {\bibinfo {title} {Practical self-consistent $gw$
  scheme for electronic structure of open-shell 3$d$-transition-metal monoxide
  anions},}\ }\bibinfo {note} {To be published elsewhere}\BibitemShut {NoStop}%
\bibitem [{\citenamefont {Faleev}\ \emph {et~al.}(2004)\citenamefont {Faleev},
  \citenamefont {van Schilfgaarde},\ and\ \citenamefont {Kotani}}]{Faleev04}%
  \BibitemOpen
  \bibfield  {author} {\bibinfo {author} {\bibfnamefont {S.~V.}\ \bibnamefont
  {Faleev}}, \bibinfo {author} {\bibfnamefont {M.}~\bibnamefont {van
  Schilfgaarde}}, \ and\ \bibinfo {author} {\bibfnamefont {T.}~\bibnamefont
  {Kotani}},\ }\href {\doibase 10.1103/PhysRevLett.93.126406} {\bibfield
  {journal} {\bibinfo  {journal} {Phys. Rev. Lett.}\ }\textbf {\bibinfo
  {volume} {93}},\ \bibinfo {pages} {126406} (\bibinfo {year}
  {2004})}\BibitemShut {NoStop}%
\bibitem [{\citenamefont {van Schilfgaarde}\ \emph {et~al.}(2006)\citenamefont
  {van Schilfgaarde}, \citenamefont {Kotani},\ and\ \citenamefont
  {Faleev}}]{vanSchilfgaarde06}%
  \BibitemOpen
  \bibfield  {author} {\bibinfo {author} {\bibfnamefont {M.}~\bibnamefont {van
  Schilfgaarde}}, \bibinfo {author} {\bibfnamefont {T.}~\bibnamefont {Kotani}},
  \ and\ \bibinfo {author} {\bibfnamefont {S.}~\bibnamefont {Faleev}},\ }\href
  {\doibase 10.1103/PhysRevLett.96.226402} {\bibfield  {journal} {\bibinfo
  {journal} {Phys. Rev. Lett.}\ }\textbf {\bibinfo {volume} {96}},\ \bibinfo
  {pages} {226402} (\bibinfo {year} {2006})}\BibitemShut {NoStop}%
\bibitem [{\citenamefont {Kotani}\ \emph {et~al.}(2007)\citenamefont {Kotani},
  \citenamefont {van Schilfgaarde},\ and\ \citenamefont {Faleev}}]{Kotani07}%
  \BibitemOpen
  \bibfield  {author} {\bibinfo {author} {\bibfnamefont {T.}~\bibnamefont
  {Kotani}}, \bibinfo {author} {\bibfnamefont {M.}~\bibnamefont {van
  Schilfgaarde}}, \ and\ \bibinfo {author} {\bibfnamefont {S.~V.}\ \bibnamefont
  {Faleev}},\ }\href {\doibase 10.1103/PhysRevB.76.165106} {\bibfield
  {journal} {\bibinfo  {journal} {Phys. Rev. B}\ }\textbf {\bibinfo {volume}
  {76}},\ \bibinfo {pages} {165106} (\bibinfo {year} {2007})}\BibitemShut
  {NoStop}%
\bibitem [{\citenamefont {Bruneval}\ and\ \citenamefont
  {Gatti}(2014)}]{Bruneval14}%
  \BibitemOpen
  \bibfield  {author} {\bibinfo {author} {\bibfnamefont {F.}~\bibnamefont
  {Bruneval}}\ and\ \bibinfo {author} {\bibfnamefont {M.}~\bibnamefont
  {Gatti}},\ }\enquote {\bibinfo {title} {Quasiparticle self-consistent gw
  method for the spectral properties of complex materials},}\ in\ \href
  {\doibase 10.1007/128_2013_460} {\emph {\bibinfo {booktitle} {First
  Principles Approaches to Spectroscopic Properties of Complex Materials}}},\
  \bibinfo {editor} {edited by\ \bibinfo {editor} {\bibfnamefont
  {C.}~\bibnamefont {Di~Valentin}}, \bibinfo {editor} {\bibfnamefont
  {S.}~\bibnamefont {Botti}}, \ and\ \bibinfo {editor} {\bibfnamefont
  {M.}~\bibnamefont {Cococcioni}}}\ (\bibinfo  {publisher} {Springer Berlin
  Heidelberg},\ \bibinfo {address} {Berlin, Heidelberg},\ \bibinfo {year}
  {2014})\ pp.\ \bibinfo {pages} {99--135}\BibitemShut {NoStop}%
\bibitem [{\citenamefont {Koval}\ \emph {et~al.}(2014)\citenamefont {Koval},
  \citenamefont {Foerster},\ and\ \citenamefont {S\'anchez-Portal}}]{Koval14}%
  \BibitemOpen
  \bibfield  {author} {\bibinfo {author} {\bibfnamefont {P.}~\bibnamefont
  {Koval}}, \bibinfo {author} {\bibfnamefont {D.}~\bibnamefont {Foerster}}, \
  and\ \bibinfo {author} {\bibfnamefont {D.}~\bibnamefont {S\'anchez-Portal}},\
  }\href {\doibase 10.1103/PhysRevB.89.155417} {\bibfield  {journal} {\bibinfo
  {journal} {Phys. Rev. B}\ }\textbf {\bibinfo {volume} {89}},\ \bibinfo
  {pages} {155417} (\bibinfo {year} {2014})}\BibitemShut {NoStop}%
\bibitem [{\citenamefont {Kaplan}\ \emph {et~al.}(2016)\citenamefont {Kaplan},
  \citenamefont {Harding}, \citenamefont {Seiler}, \citenamefont {Weigend},
  \citenamefont {Evers},\ and\ \citenamefont {van Setten}}]{Kaplan16}%
  \BibitemOpen
  \bibfield  {author} {\bibinfo {author} {\bibfnamefont {F.}~\bibnamefont
  {Kaplan}}, \bibinfo {author} {\bibfnamefont {M.~E.}\ \bibnamefont {Harding}},
  \bibinfo {author} {\bibfnamefont {C.}~\bibnamefont {Seiler}}, \bibinfo
  {author} {\bibfnamefont {F.}~\bibnamefont {Weigend}}, \bibinfo {author}
  {\bibfnamefont {F.}~\bibnamefont {Evers}}, \ and\ \bibinfo {author}
  {\bibfnamefont {M.~J.}\ \bibnamefont {van Setten}},\ }\href {\doibase
  10.1021/acs.jctc.5b01238} {\bibfield  {journal} {\bibinfo  {journal} {Journal
  of Chemical Theory and Computation}\ }\textbf {\bibinfo {volume} {12}},\
  \bibinfo {pages} {2528} (\bibinfo {year} {2016})},\ \bibinfo {note} {pMID:
  27168352},\ \Eprint
  {http://arxiv.org/abs/https://doi.org/10.1021/acs.jctc.5b01238}
  {https://doi.org/10.1021/acs.jctc.5b01238} \BibitemShut {NoStop}%
\bibitem [{\citenamefont {Caruso}\ \emph {et~al.}(2016)\citenamefont {Caruso},
  \citenamefont {Dauth}, \citenamefont {van Setten},\ and\ \citenamefont
  {Rinke}}]{Caruso16}%
  \BibitemOpen
  \bibfield  {author} {\bibinfo {author} {\bibfnamefont {F.}~\bibnamefont
  {Caruso}}, \bibinfo {author} {\bibfnamefont {M.}~\bibnamefont {Dauth}},
  \bibinfo {author} {\bibfnamefont {M.~J.}\ \bibnamefont {van Setten}}, \ and\
  \bibinfo {author} {\bibfnamefont {P.}~\bibnamefont {Rinke}},\ }\href
  {\doibase 10.1021/acs.jctc.6b00774} {\bibfield  {journal} {\bibinfo
  {journal} {Journal of Chemical Theory and Computation}\ }\textbf {\bibinfo
  {volume} {12}},\ \bibinfo {pages} {5076} (\bibinfo {year} {2016})},\ \bibinfo
  {note} {pMID: 27631585},\ \Eprint
  {http://arxiv.org/abs/https://doi.org/10.1021/acs.jctc.6b00774}
  {https://doi.org/10.1021/acs.jctc.6b00774} \BibitemShut {NoStop}%
\bibitem [{\citenamefont {van Setten}\ \emph {et~al.}(2018)\citenamefont {van
  Setten}, \citenamefont {Costa}, \citenamefont {Vi\~{n}es},\ and\
  \citenamefont {Illas}}]{vanSetten18}%
  \BibitemOpen
  \bibfield  {author} {\bibinfo {author} {\bibfnamefont {M.~J.}\ \bibnamefont
  {van Setten}}, \bibinfo {author} {\bibfnamefont {R.}~\bibnamefont {Costa}},
  \bibinfo {author} {\bibfnamefont {F.}~\bibnamefont {Vi\~{n}es}}, \ and\
  \bibinfo {author} {\bibfnamefont {F.}~\bibnamefont {Illas}},\ }\href
  {\doibase 10.1021/acs.jctc.7b01192} {\bibfield  {journal} {\bibinfo
  {journal} {Journal of Chemical Theory and Computation}\ }\textbf {\bibinfo
  {volume} {14}},\ \bibinfo {pages} {877} (\bibinfo {year} {2018})},\ \bibinfo
  {note} {pMID: 29320628},\ \Eprint
  {http://arxiv.org/abs/https://doi.org/10.1021/acs.jctc.7b01192}
  {https://doi.org/10.1021/acs.jctc.7b01192} \BibitemShut {NoStop}%
\bibitem [{\citenamefont {Ben}\ \emph {et~al.}(2020)\citenamefont {Ben},
  \citenamefont {Yang}, \citenamefont {Li}, \citenamefont {Jornada},
  \citenamefont {Louie},\ and\ \citenamefont {Deslippe}}]{Ben20}%
  \BibitemOpen
  \bibfield  {author} {\bibinfo {author} {\bibfnamefont {M.~D.}\ \bibnamefont
  {Ben}}, \bibinfo {author} {\bibfnamefont {C.}~\bibnamefont {Yang}}, \bibinfo
  {author} {\bibfnamefont {Z.}~\bibnamefont {Li}}, \bibinfo {author}
  {\bibfnamefont {F.~H.~d.}\ \bibnamefont {Jornada}}, \bibinfo {author}
  {\bibfnamefont {S.~G.}\ \bibnamefont {Louie}}, \ and\ \bibinfo {author}
  {\bibfnamefont {J.}~\bibnamefont {Deslippe}},\ }in\ \href {\doibase
  10.1109/SC41405.2020.00008} {\emph {\bibinfo {booktitle} {SC20: International
  Conference for High Performance Computing, Networking, Storage and
  Analysis}}}\ (\bibinfo {year} {2020})\ pp.\ \bibinfo {pages}
  {1--11}\BibitemShut {NoStop}%
\bibitem [{\citenamefont {Barca}\ \emph {et~al.}(2021)\citenamefont {Barca},
  \citenamefont {Alkan}, \citenamefont {Galvez-Vallejo}, \citenamefont {Poole},
  \citenamefont {Rendell},\ and\ \citenamefont {Gordon}}]{Barca21}%
  \BibitemOpen
  \bibfield  {author} {\bibinfo {author} {\bibfnamefont {G.~M.~J.}\
  \bibnamefont {Barca}}, \bibinfo {author} {\bibfnamefont {M.}~\bibnamefont
  {Alkan}}, \bibinfo {author} {\bibfnamefont {J.~L.}\ \bibnamefont
  {Galvez-Vallejo}}, \bibinfo {author} {\bibfnamefont {D.~L.}\ \bibnamefont
  {Poole}}, \bibinfo {author} {\bibfnamefont {A.~P.}\ \bibnamefont {Rendell}},
  \ and\ \bibinfo {author} {\bibfnamefont {M.~S.}\ \bibnamefont {Gordon}},\
  }\href {\doibase 10.1021/acs.jctc.1c00720} {\bibfield  {journal} {\bibinfo
  {journal} {Journal of Chemical Theory and Computation}\ }\textbf {\bibinfo
  {volume} {17}},\ \bibinfo {pages} {7486} (\bibinfo {year} {2021})},\ \bibinfo
  {note} {pMID: 34780186},\ \Eprint
  {http://arxiv.org/abs/https://doi.org/10.1021/acs.jctc.1c00720}
  {https://doi.org/10.1021/acs.jctc.1c00720} \BibitemShut {NoStop}%
\bibitem [{\citenamefont {Yu}\ and\ \citenamefont {Govoni}(2022)}]{Yu22}%
  \BibitemOpen
  \bibfield  {author} {\bibinfo {author} {\bibfnamefont {V.~W.-z.}\
  \bibnamefont {Yu}}\ and\ \bibinfo {author} {\bibfnamefont {M.}~\bibnamefont
  {Govoni}},\ }\href {\doibase 10.1021/acs.jctc.2c00241} {\bibfield  {journal}
  {\bibinfo  {journal} {Journal of Chemical Theory and Computation}\ }\textbf
  {\bibinfo {volume} {18}},\ \bibinfo {pages} {4690} (\bibinfo {year}
  {2022})},\ \bibinfo {note} {pMID: 35913080},\ \Eprint
  {http://arxiv.org/abs/https://doi.org/10.1021/acs.jctc.2c00241}
  {https://doi.org/10.1021/acs.jctc.2c00241} \BibitemShut {NoStop}%
\bibitem [{\citenamefont {Needham}\ \emph
  {et~al.}(2016{\natexlab{a}})\citenamefont {Needham}, \citenamefont
  {G\"{o}tz},\ and\ \citenamefont {Walker}}]{Walker16a}%
  \BibitemOpen
  \bibfield  {author} {\bibinfo {author} {\bibfnamefont {P.}~\bibnamefont
  {Needham}}, \bibinfo {author} {\bibfnamefont {A.~W.}\ \bibnamefont
  {G\"{o}tz}}, \ and\ \bibinfo {author} {\bibfnamefont {R.~C.}\ \bibnamefont
  {Walker}},\ }\enquote {\bibinfo {title} {Why graphics processing units},}\
  in\ \href {\doibase https://doi.org/10.1002/9781118670712.ch1} {\emph
  {\bibinfo {booktitle} {Electronic Structure Calculations on Graphics
  Processing Units}}}\ (\bibinfo  {publisher} {John Wiley \& Sons, Ltd},\
  \bibinfo {year} {2016})\ Chap.~\bibinfo {chapter} {1}, pp.\ \bibinfo {pages}
  {1--22},\ \Eprint
  {http://arxiv.org/abs/https://onlinelibrary.wiley.com/doi/pdf/10.1002/9781118670712.ch1}
  {https://onlinelibrary.wiley.com/doi/pdf/10.1002/9781118670712.ch1}
  \BibitemShut {NoStop}%
\bibitem [{\citenamefont {Needham}\ \emph
  {et~al.}(2016{\natexlab{b}})\citenamefont {Needham}, \citenamefont
  {G\"{o}tz},\ and\ \citenamefont {Walker}}]{Walker16b}%
  \BibitemOpen
  \bibfield  {author} {\bibinfo {author} {\bibfnamefont {P.}~\bibnamefont
  {Needham}}, \bibinfo {author} {\bibfnamefont {A.~W.}\ \bibnamefont
  {G\"{o}tz}}, \ and\ \bibinfo {author} {\bibfnamefont {R.~C.}\ \bibnamefont
  {Walker}},\ }\enquote {\bibinfo {title} {Gpus: Hardware to software},}\ in\
  \href {\doibase https://doi.org/10.1002/9781118670712.ch2} {\emph {\bibinfo
  {booktitle} {Electronic Structure Calculations on Graphics Processing
  Units}}}\ (\bibinfo  {publisher} {John Wiley \& Sons, Ltd},\ \bibinfo {year}
  {2016})\ Chap.~\bibinfo {chapter} {2}, pp.\ \bibinfo {pages} {23--38},\
  \Eprint
  {http://arxiv.org/abs/https://onlinelibrary.wiley.com/doi/pdf/10.1002/9781118670712.ch2}
  {https://onlinelibrary.wiley.com/doi/pdf/10.1002/9781118670712.ch2}
  \BibitemShut {NoStop}%
\bibitem [{\citenamefont {G\"{o}tz}(2016)}]{Gotz16}%
  \BibitemOpen
  \bibfield  {author} {\bibinfo {author} {\bibfnamefont {A.~W.}\ \bibnamefont
  {G\"{o}tz}},\ }\enquote {\bibinfo {title} {Overview of electronic structure
  methods},}\ in\ \href {\doibase https://doi.org/10.1002/9781118670712.ch3}
  {\emph {\bibinfo {booktitle} {Electronic Structure Calculations on Graphics
  Processing Units}}}\ (\bibinfo  {publisher} {John Wiley \& Sons, Ltd},\
  \bibinfo {year} {2016})\ Chap.~\bibinfo {chapter} {3}, pp.\ \bibinfo {pages}
  {39--66},\ \Eprint
  {http://arxiv.org/abs/https://onlinelibrary.wiley.com/doi/pdf/10.1002/9781118670712.ch3}
  {https://onlinelibrary.wiley.com/doi/pdf/10.1002/9781118670712.ch3}
  \BibitemShut {NoStop}%
\bibitem [{\citenamefont {DePrince}\ and\ \citenamefont
  {Hammond}(2011)}]{DePrince11}%
  \BibitemOpen
  \bibfield  {author} {\bibinfo {author} {\bibfnamefont {A.~E.~I.}\
  \bibnamefont {DePrince}}\ and\ \bibinfo {author} {\bibfnamefont {J.~R.}\
  \bibnamefont {Hammond}},\ }\href {\doibase 10.1021/ct100584w} {\bibfield
  {journal} {\bibinfo  {journal} {Journal of Chemical Theory and Computation}\
  }\textbf {\bibinfo {volume} {7}},\ \bibinfo {pages} {1287} (\bibinfo {year}
  {2011})},\ \bibinfo {note} {pMID: 26610123},\ \Eprint
  {http://arxiv.org/abs/https://doi.org/10.1021/ct100584w}
  {https://doi.org/10.1021/ct100584w} \BibitemShut {NoStop}%
\bibitem [{\citenamefont {Isborn}\ \emph {et~al.}(2011)\citenamefont {Isborn},
  \citenamefont {Luehr}, \citenamefont {Ufimtsev},\ and\ \citenamefont
  {Mart\'{i}nez}}]{Isborn11}%
  \BibitemOpen
  \bibfield  {author} {\bibinfo {author} {\bibfnamefont {C.~M.}\ \bibnamefont
  {Isborn}}, \bibinfo {author} {\bibfnamefont {N.}~\bibnamefont {Luehr}},
  \bibinfo {author} {\bibfnamefont {I.~S.}\ \bibnamefont {Ufimtsev}}, \ and\
  \bibinfo {author} {\bibfnamefont {T.~J.}\ \bibnamefont {Mart\'{i}nez}},\
  }\href {\doibase 10.1021/ct200030k} {\bibfield  {journal} {\bibinfo
  {journal} {Journal of Chemical Theory and Computation}\ }\textbf {\bibinfo
  {volume} {7}},\ \bibinfo {pages} {1814} (\bibinfo {year} {2011})},\ \bibinfo
  {note} {pMID: 21687784},\ \Eprint
  {http://arxiv.org/abs/https://doi.org/10.1021/ct200030k}
  {https://doi.org/10.1021/ct200030k} \BibitemShut {NoStop}%
\bibitem [{\citenamefont {Gordon}\ \emph {et~al.}(2020)\citenamefont {Gordon},
  \citenamefont {Barca}, \citenamefont {Leang}, \citenamefont {Poole},
  \citenamefont {Rendell}, \citenamefont {Galvez~Vallejo},\ and\ \citenamefont
  {Westheimer}}]{Gordon20a}%
  \BibitemOpen
  \bibfield  {author} {\bibinfo {author} {\bibfnamefont {M.~S.}\ \bibnamefont
  {Gordon}}, \bibinfo {author} {\bibfnamefont {G.}~\bibnamefont {Barca}},
  \bibinfo {author} {\bibfnamefont {S.~S.}\ \bibnamefont {Leang}}, \bibinfo
  {author} {\bibfnamefont {D.}~\bibnamefont {Poole}}, \bibinfo {author}
  {\bibfnamefont {A.~P.}\ \bibnamefont {Rendell}}, \bibinfo {author}
  {\bibfnamefont {J.~L.}\ \bibnamefont {Galvez~Vallejo}}, \ and\ \bibinfo
  {author} {\bibfnamefont {B.}~\bibnamefont {Westheimer}},\ }\href {\doibase
  10.1021/acs.jpca.0c02249} {\bibfield  {journal} {\bibinfo  {journal} {The
  Journal of Physical Chemistry A}\ }\textbf {\bibinfo {volume} {124}},\
  \bibinfo {pages} {4557} (\bibinfo {year} {2020})},\ \bibinfo {note} {pMID:
  32379450},\ \Eprint
  {http://arxiv.org/abs/https://doi.org/10.1021/acs.jpca.0c02249}
  {https://doi.org/10.1021/acs.jpca.0c02249} \BibitemShut {NoStop}%
\bibitem [{\citenamefont {Gordon}\ and\ \citenamefont
  {Windus}(2020)}]{Gordon20b}%
  \BibitemOpen
  \bibfield  {author} {\bibinfo {author} {\bibfnamefont {M.~S.}\ \bibnamefont
  {Gordon}}\ and\ \bibinfo {author} {\bibfnamefont {T.~L.}\ \bibnamefont
  {Windus}},\ }\href {\doibase 10.1021/acs.chemrev.0c00700} {\bibfield
  {journal} {\bibinfo  {journal} {Chemical Reviews}\ }\textbf {\bibinfo
  {volume} {120}},\ \bibinfo {pages} {9015} (\bibinfo {year} {2020})},\
  \bibinfo {note} {pMID: 32900196},\ \Eprint
  {http://arxiv.org/abs/https://doi.org/10.1021/acs.chemrev.0c00700}
  {https://doi.org/10.1021/acs.chemrev.0c00700} \BibitemShut {NoStop}%
\bibitem [{\citenamefont {Calvin}\ \emph {et~al.}(2021)\citenamefont {Calvin},
  \citenamefont {Peng}, \citenamefont {Rishi}, \citenamefont {Kumar},\ and\
  \citenamefont {Valeev}}]{Calvin21}%
  \BibitemOpen
  \bibfield  {author} {\bibinfo {author} {\bibfnamefont {J.~A.}\ \bibnamefont
  {Calvin}}, \bibinfo {author} {\bibfnamefont {C.}~\bibnamefont {Peng}},
  \bibinfo {author} {\bibfnamefont {V.}~\bibnamefont {Rishi}}, \bibinfo
  {author} {\bibfnamefont {A.}~\bibnamefont {Kumar}}, \ and\ \bibinfo {author}
  {\bibfnamefont {E.~F.}\ \bibnamefont {Valeev}},\ }\href {\doibase
  10.1021/acs.chemrev.0c00006} {\bibfield  {journal} {\bibinfo  {journal}
  {Chemical Reviews}\ }\textbf {\bibinfo {volume} {121}},\ \bibinfo {pages}
  {1203} (\bibinfo {year} {2021})},\ \bibinfo {note} {pMID: 33305957},\ \Eprint
  {http://arxiv.org/abs/https://doi.org/10.1021/acs.chemrev.0c00006}
  {https://doi.org/10.1021/acs.chemrev.0c00006} \BibitemShut {NoStop}%
\bibitem [{CUD()}]{CUDA}%
  \BibitemOpen
  \href@noop {} {\enquote {\bibinfo {title} {Nvidia cuda homepage},}\ }\bibinfo
  {howpublished} {\url{https://developer.nvidia.com/cuda-zone}},\ \bibinfo
  {note} {accessed June 8, 2023}\BibitemShut {NoStop}%
\bibitem [{Ope({\natexlab{a}})}]{OpenACC}%
  \BibitemOpen
  \href@noop {} {\enquote {\bibinfo {title} {Openacc homepage},}\ }\bibinfo
  {howpublished} {\url{https://www.openacc.org/}} ({\natexlab{a}}),\ \bibinfo
  {note} {accessed June 8, 2023}\BibitemShut {NoStop}%
\bibitem [{Dro()}]{Dropbox}%
  \BibitemOpen
  \href@noop {} {\enquote {\bibinfo {title} {Dropbox blog},}\ }\bibinfo
  {howpublished}
  {\url{https://blog.dropbox.com/topics/company/thank-you--guido}},\ \bibinfo
  {note} {accessed June 8, 2023}\BibitemShut {NoStop}%
\bibitem [{\citenamefont {Eriksen}(2017)}]{Eriksen17}%
  \BibitemOpen
  \bibfield  {author} {\bibinfo {author} {\bibfnamefont {J.~J.}\ \bibnamefont
  {Eriksen}},\ }\href {\doibase 10.1080/00268976.2016.1271155} {\bibfield
  {journal} {\bibinfo  {journal} {Molecular Physics}\ }\textbf {\bibinfo
  {volume} {115}},\ \bibinfo {pages} {2086} (\bibinfo {year} {2017})},\ \Eprint
  {http://arxiv.org/abs/https://doi.org/10.1080/00268976.2016.1271155}
  {https://doi.org/10.1080/00268976.2016.1271155} \BibitemShut {NoStop}%
\bibitem [{\citenamefont {Bykov}\ and\ \citenamefont
  {Kjaergaard}(2017)}]{Bykov17}%
  \BibitemOpen
  \bibfield  {author} {\bibinfo {author} {\bibfnamefont {D.}~\bibnamefont
  {Bykov}}\ and\ \bibinfo {author} {\bibfnamefont {T.}~\bibnamefont
  {Kjaergaard}},\ }\href {\doibase https://doi.org/10.1002/jcc.24678}
  {\bibfield  {journal} {\bibinfo  {journal} {Journal of Computational
  Chemistry}\ }\textbf {\bibinfo {volume} {38}},\ \bibinfo {pages} {228}
  (\bibinfo {year} {2017})},\ \Eprint
  {http://arxiv.org/abs/https://onlinelibrary.wiley.com/doi/pdf/10.1002/jcc.24678}
  {https://onlinelibrary.wiley.com/doi/pdf/10.1002/jcc.24678} \BibitemShut
  {NoStop}%
\bibitem [{\citenamefont {Vergara~Larrea}\ \emph {et~al.}(2020)\citenamefont
  {Vergara~Larrea}, \citenamefont {Budiardja}, \citenamefont {Gayatri},
  \citenamefont {Daley}, \citenamefont {Hernandez},\ and\ \citenamefont
  {Joubert}}]{VergaraLarrea20}%
  \BibitemOpen
  \bibfield  {author} {\bibinfo {author} {\bibfnamefont {V.~G.}\ \bibnamefont
  {Vergara~Larrea}}, \bibinfo {author} {\bibfnamefont {R.~D.}\ \bibnamefont
  {Budiardja}}, \bibinfo {author} {\bibfnamefont {R.}~\bibnamefont {Gayatri}},
  \bibinfo {author} {\bibfnamefont {C.}~\bibnamefont {Daley}}, \bibinfo
  {author} {\bibfnamefont {O.}~\bibnamefont {Hernandez}}, \ and\ \bibinfo
  {author} {\bibfnamefont {W.}~\bibnamefont {Joubert}},\ }\href {\doibase
  https://doi.org/10.1002/cpe.5780} {\bibfield  {journal} {\bibinfo  {journal}
  {Concurrency and Computation: Practice and Experience}\ }\textbf {\bibinfo
  {volume} {32}},\ \bibinfo {pages} {e5780} (\bibinfo {year} {2020})},\ \Eprint
  {http://arxiv.org/abs/https://onlinelibrary.wiley.com/doi/pdf/10.1002/cpe.5780}
  {https://onlinelibrary.wiley.com/doi/pdf/10.1002/cpe.5780} \BibitemShut
  {NoStop}%
\bibitem [{\citenamefont {Smith}\ \emph {et~al.}(2022)\citenamefont {Smith},
  \citenamefont {Tamerus},\ and\ \citenamefont {Hasnip}}]{Smith22}%
  \BibitemOpen
  \bibfield  {author} {\bibinfo {author} {\bibfnamefont {M.}~\bibnamefont
  {Smith}}, \bibinfo {author} {\bibfnamefont {A.}~\bibnamefont {Tamerus}}, \
  and\ \bibinfo {author} {\bibfnamefont {P.}~\bibnamefont {Hasnip}},\ }\href
  {\doibase 10.1109/MCSE.2022.3141714} {\bibfield  {journal} {\bibinfo
  {journal} {Computing in Science \& Engineering}\ }\textbf {\bibinfo {volume}
  {24}},\ \bibinfo {pages} {46} (\bibinfo {year} {2022})}\BibitemShut {NoStop}%
\bibitem [{\citenamefont {Dang}\ \emph {et~al.}(2022)\citenamefont {Dang},
  \citenamefont {Wilson},\ and\ \citenamefont {Zimmerman}}]{Dang22}%
  \BibitemOpen
  \bibfield  {author} {\bibinfo {author} {\bibfnamefont {D.-K.}\ \bibnamefont
  {Dang}}, \bibinfo {author} {\bibfnamefont {L.~W.}\ \bibnamefont {Wilson}}, \
  and\ \bibinfo {author} {\bibfnamefont {P.~M.}\ \bibnamefont {Zimmerman}},\
  }\href {\doibase https://doi.org/10.1002/jcc.26968} {\bibfield  {journal}
  {\bibinfo  {journal} {Journal of Computational Chemistry}\ }\textbf {\bibinfo
  {volume} {43}},\ \bibinfo {pages} {1680} (\bibinfo {year} {2022})},\ \Eprint
  {http://arxiv.org/abs/https://onlinelibrary.wiley.com/doi/pdf/10.1002/jcc.26968}
  {https://onlinelibrary.wiley.com/doi/pdf/10.1002/jcc.26968} \BibitemShut
  {NoStop}%
\bibitem [{\citenamefont {Maintz}\ and\ \citenamefont
  {Wetzstein}(2018)}]{Maintz18a}%
  \BibitemOpen
  \bibfield  {author} {\bibinfo {author} {\bibfnamefont {S.}~\bibnamefont
  {Maintz}}\ and\ \bibinfo {author} {\bibfnamefont {M.}~\bibnamefont
  {Wetzstein}}\ }(\bibinfo {year} {2018})\BibitemShut {NoStop}%
\bibitem [{\citenamefont {Maintz}()}]{Maintz18b}%
  \BibitemOpen
  \bibfield  {author} {\bibinfo {author} {\bibfnamefont {S.}~\bibnamefont
  {Maintz}},\ }\href@noop {} {\enquote {\bibinfo {title} {Porting vasp to gpus
  with openacc},}\ }\bibinfo {howpublished}
  {\url{https://www.nvidia.com/en-us/on-demand/session/gtceurope2018-e8367}},\
  \bibinfo {note} {accessed June 8, 2023}\BibitemShut {NoStop}%
\bibitem [{\citenamefont {Blase}\ \emph {et~al.}(2011)\citenamefont {Blase},
  \citenamefont {Attaccalite},\ and\ \citenamefont {Olevano}}]{Blase11}%
  \BibitemOpen
  \bibfield  {author} {\bibinfo {author} {\bibfnamefont {X.}~\bibnamefont
  {Blase}}, \bibinfo {author} {\bibfnamefont {C.}~\bibnamefont {Attaccalite}},
  \ and\ \bibinfo {author} {\bibfnamefont {V.}~\bibnamefont {Olevano}},\ }\href
  {\doibase 10.1103/PhysRevB.83.115103} {\bibfield  {journal} {\bibinfo
  {journal} {Phys. Rev. B}\ }\textbf {\bibinfo {volume} {83}},\ \bibinfo
  {pages} {115103} (\bibinfo {year} {2011})}\BibitemShut {NoStop}%
\bibitem [{\citenamefont {Ren}\ \emph {et~al.}(2012)\citenamefont {Ren},
  \citenamefont {Rinke}, \citenamefont {Blum}, \citenamefont {Wieferink},
  \citenamefont {Tkatchenko}, \citenamefont {Sanfilippo}, \citenamefont
  {Reuter},\ and\ \citenamefont {Scheffler}}]{Ren12}%
  \BibitemOpen
  \bibfield  {author} {\bibinfo {author} {\bibfnamefont {X.}~\bibnamefont
  {Ren}}, \bibinfo {author} {\bibfnamefont {P.}~\bibnamefont {Rinke}}, \bibinfo
  {author} {\bibfnamefont {V.}~\bibnamefont {Blum}}, \bibinfo {author}
  {\bibfnamefont {J.}~\bibnamefont {Wieferink}}, \bibinfo {author}
  {\bibfnamefont {A.}~\bibnamefont {Tkatchenko}}, \bibinfo {author}
  {\bibfnamefont {A.}~\bibnamefont {Sanfilippo}}, \bibinfo {author}
  {\bibfnamefont {K.}~\bibnamefont {Reuter}}, \ and\ \bibinfo {author}
  {\bibfnamefont {M.}~\bibnamefont {Scheffler}},\ }\href
  {http://stacks.iop.org/1367-2630/14/i=5/a=053020} {\bibfield  {journal}
  {\bibinfo  {journal} {New Journal of Physics}\ }\textbf {\bibinfo {volume}
  {14}},\ \bibinfo {pages} {053020} (\bibinfo {year} {2012})}\BibitemShut
  {NoStop}%
\bibitem [{\citenamefont {van Setten}\ \emph {et~al.}(2013)\citenamefont {van
  Setten}, \citenamefont {Weigend},\ and\ \citenamefont {Evers}}]{vanSetten13}%
  \BibitemOpen
  \bibfield  {author} {\bibinfo {author} {\bibfnamefont {M.~J.}\ \bibnamefont
  {van Setten}}, \bibinfo {author} {\bibfnamefont {F.}~\bibnamefont {Weigend}},
  \ and\ \bibinfo {author} {\bibfnamefont {F.}~\bibnamefont {Evers}},\ }\href
  {\doibase 10.1021/ct300648t} {\bibfield  {journal} {\bibinfo  {journal}
  {Journal of Chemical Theory and Computation}\ }\textbf {\bibinfo {volume}
  {9}},\ \bibinfo {pages} {232} (\bibinfo {year} {2013})},\ \bibinfo {note}
  {pMID: 26589026},\ \Eprint
  {http://arxiv.org/abs/https://doi.org/10.1021/ct300648t}
  {https://doi.org/10.1021/ct300648t} \BibitemShut {NoStop}%
\bibitem [{\citenamefont {Wilhelm}\ \emph {et~al.}(2016)\citenamefont
  {Wilhelm}, \citenamefont {Del~Ben},\ and\ \citenamefont
  {Hutter}}]{Wilhelm16}%
  \BibitemOpen
  \bibfield  {author} {\bibinfo {author} {\bibfnamefont {J.}~\bibnamefont
  {Wilhelm}}, \bibinfo {author} {\bibfnamefont {M.}~\bibnamefont {Del~Ben}}, \
  and\ \bibinfo {author} {\bibfnamefont {J.}~\bibnamefont {Hutter}},\ }\href
  {\doibase 10.1021/acs.jctc.6b00380} {\bibfield  {journal} {\bibinfo
  {journal} {Journal of Chemical Theory and Computation}\ }\textbf {\bibinfo
  {volume} {12}},\ \bibinfo {pages} {3623} (\bibinfo {year} {2016})},\ \bibinfo
  {note} {pMID: 27348184},\ \Eprint
  {http://arxiv.org/abs/https://doi.org/10.1021/acs.jctc.6b00380}
  {https://doi.org/10.1021/acs.jctc.6b00380} \BibitemShut {NoStop}%
\bibitem [{\citenamefont {Bruneval}\ \emph {et~al.}(2016)\citenamefont
  {Bruneval}, \citenamefont {Rangel}, \citenamefont {Hamed}, \citenamefont
  {Shao}, \citenamefont {Yang},\ and\ \citenamefont {Neaton}}]{Bruneval16}%
  \BibitemOpen
  \bibfield  {author} {\bibinfo {author} {\bibfnamefont {F.}~\bibnamefont
  {Bruneval}}, \bibinfo {author} {\bibfnamefont {T.}~\bibnamefont {Rangel}},
  \bibinfo {author} {\bibfnamefont {S.~M.}\ \bibnamefont {Hamed}}, \bibinfo
  {author} {\bibfnamefont {M.}~\bibnamefont {Shao}}, \bibinfo {author}
  {\bibfnamefont {C.}~\bibnamefont {Yang}}, \ and\ \bibinfo {author}
  {\bibfnamefont {J.~B.}\ \bibnamefont {Neaton}},\ }\href {\doibase
  https://doi.org/10.1016/j.cpc.2016.06.019} {\bibfield  {journal} {\bibinfo
  {journal} {Computer Physics Communications}\ }\textbf {\bibinfo {volume}
  {208}},\ \bibinfo {pages} {149 } (\bibinfo {year} {2016})}\BibitemShut
  {NoStop}%
\bibitem [{\citenamefont {Golze}\ \emph {et~al.}(2019)\citenamefont {Golze},
  \citenamefont {Dvorak},\ and\ \citenamefont {Rinke}}]{Golze19}%
  \BibitemOpen
  \bibfield  {author} {\bibinfo {author} {\bibfnamefont {D.}~\bibnamefont
  {Golze}}, \bibinfo {author} {\bibfnamefont {M.}~\bibnamefont {Dvorak}}, \
  and\ \bibinfo {author} {\bibfnamefont {P.}~\bibnamefont {Rinke}},\ }\href
  {\doibase 10.3389/fchem.2019.00377} {\bibfield  {journal} {\bibinfo
  {journal} {Frontiers in Chemistry}\ }\textbf {\bibinfo {volume} {7}},\
  \bibinfo {pages} {377} (\bibinfo {year} {2019})}\BibitemShut {NoStop}%
\bibitem [{\citenamefont {Bruneval}\ \emph {et~al.}(2021)\citenamefont
  {Bruneval}, \citenamefont {Dattani},\ and\ \citenamefont {van
  Setten}}]{Bruneval21}%
  \BibitemOpen
  \bibfield  {author} {\bibinfo {author} {\bibfnamefont {F.}~\bibnamefont
  {Bruneval}}, \bibinfo {author} {\bibfnamefont {N.}~\bibnamefont {Dattani}}, \
  and\ \bibinfo {author} {\bibfnamefont {M.~J.}\ \bibnamefont {van Setten}},\
  }\href {\doibase 10.3389/fchem.2021.749779} {\bibfield  {journal} {\bibinfo
  {journal} {Frontiers in Chemistry}\ }\textbf {\bibinfo {volume} {9}}
  (\bibinfo {year} {2021}),\ 10.3389/fchem.2021.749779}\BibitemShut {NoStop}%
\bibitem [{\citenamefont {Bruneval}\ \emph {et~al.}(2006)\citenamefont
  {Bruneval}, \citenamefont {Vast},\ and\ \citenamefont
  {Reining}}]{Bruneval06}%
  \BibitemOpen
  \bibfield  {author} {\bibinfo {author} {\bibfnamefont {F.}~\bibnamefont
  {Bruneval}}, \bibinfo {author} {\bibfnamefont {N.}~\bibnamefont {Vast}}, \
  and\ \bibinfo {author} {\bibfnamefont {L.}~\bibnamefont {Reining}},\ }\href
  {\doibase 10.1103/PhysRevB.74.045102} {\bibfield  {journal} {\bibinfo
  {journal} {Phys. Rev. B}\ }\textbf {\bibinfo {volume} {74}},\ \bibinfo
  {pages} {045102} (\bibinfo {year} {2006})}\BibitemShut {NoStop}%
\bibitem [{\citenamefont {Bruneval}(2012)}]{Bruneval12}%
  \BibitemOpen
  \bibfield  {author} {\bibinfo {author} {\bibfnamefont {F.}~\bibnamefont
  {Bruneval}},\ }\href {\doibase 10.1063/1.4718428} {\bibfield  {journal}
  {\bibinfo  {journal} {The Journal of Chemical Physics}\ }\textbf {\bibinfo
  {volume} {136}},\ \bibinfo {pages} {194107} (\bibinfo {year} {2012})},\
  \Eprint {http://arxiv.org/abs/https://doi.org/10.1063/1.4718428}
  {https://doi.org/10.1063/1.4718428} \BibitemShut {NoStop}%
\bibitem [{\citenamefont {CrawfordGroup}()}]{ProgrammingChemistry}%
  \BibitemOpen
  \bibfield  {author} {\bibinfo {author} {\bibnamefont {CrawfordGroup}},\
  }\href@noop {} {\enquote {\bibinfo {title} {C++ programming tutorial in
  chemistry},}\ }\bibinfo {howpublished}
  {\url{https://github.com/CrawfordGroup/ProgrammingProjects/}},\ \bibinfo
  {note} {accessed November 19, 2023}\BibitemShut {NoStop}%
\bibitem [{Ope({\natexlab{b}})}]{OpenMP}%
  \BibitemOpen
  \href@noop {} {\enquote {\bibinfo {title} {Openmp homepage},}\ }\bibinfo
  {howpublished} {\url{https://www.openmp.org/}} ({\natexlab{b}}),\ \bibinfo
  {note} {accessed June 8, 2023}\BibitemShut {NoStop}%
\bibitem [{\citenamefont {Pham}\ \emph {et~al.}(2023)\citenamefont {Pham},
  \citenamefont {Alkan},\ and\ \citenamefont {Gordon}}]{Pham23}%
  \BibitemOpen
  \bibfield  {author} {\bibinfo {author} {\bibfnamefont {B.~Q.}\ \bibnamefont
  {Pham}}, \bibinfo {author} {\bibfnamefont {M.}~\bibnamefont {Alkan}}, \ and\
  \bibinfo {author} {\bibfnamefont {M.~S.}\ \bibnamefont {Gordon}},\ }\href
  {\doibase 10.1021/acs.jctc.2c01137} {\bibfield  {journal} {\bibinfo
  {journal} {Journal of Chemical Theory and Computation}\ }\textbf {\bibinfo
  {volume} {19}},\ \bibinfo {pages} {2213} (\bibinfo {year} {2023})},\ \bibinfo
  {note} {pMID: 37011288},\ \Eprint
  {http://arxiv.org/abs/https://doi.org/10.1021/acs.jctc.2c01137}
  {https://doi.org/10.1021/acs.jctc.2c01137} \BibitemShut {NoStop}%
\bibitem [{\citenamefont {Byun}()}]{GitHub}%
  \BibitemOpen
  \bibfield  {author} {\bibinfo {author} {\bibfnamefont {Y.-M.}\ \bibnamefont
  {Byun}},\ }\href@noop {} {\enquote {\bibinfo {title} {Molgw 1.f with openmp
  and openacc (single gpu)},}\ }\bibinfo {howpublished}
  {\url{https://github.com/ymbyun/molgw-1.F-openmp-openacc-single-gpu}},\
  \bibinfo {note} {accessed June 8, 2023}\BibitemShut {NoStop}%
\bibitem [{\citenamefont {Bonati}\ \emph {et~al.}(2015)\citenamefont {Bonati},
  \citenamefont {Calore}, \citenamefont {Coscetti}, \citenamefont {D'Elia},
  \citenamefont {Mesiti}, \citenamefont {Negro}, \citenamefont {Schifano},\
  and\ \citenamefont {Tripiccione}}]{Bonati15}%
  \BibitemOpen
  \bibfield  {author} {\bibinfo {author} {\bibfnamefont {C.}~\bibnamefont
  {Bonati}}, \bibinfo {author} {\bibfnamefont {E.}~\bibnamefont {Calore}},
  \bibinfo {author} {\bibfnamefont {S.}~\bibnamefont {Coscetti}}, \bibinfo
  {author} {\bibfnamefont {M.}~\bibnamefont {D'Elia}}, \bibinfo {author}
  {\bibfnamefont {M.}~\bibnamefont {Mesiti}}, \bibinfo {author} {\bibfnamefont
  {F.}~\bibnamefont {Negro}}, \bibinfo {author} {\bibfnamefont {S.~F.}\
  \bibnamefont {Schifano}}, \ and\ \bibinfo {author} {\bibfnamefont
  {R.}~\bibnamefont {Tripiccione}},\ }in\ \href@noop {} {\emph {\bibinfo
  {booktitle} {Proceedings of the 2015 International Workshop on Software
  Engineering for High Performance Computing in Science}}},\ \bibinfo {series
  and number} {SE4HPCS '15}\ (\bibinfo  {publisher} {IEEE Press},\ \bibinfo
  {year} {2015})\ p.\ \bibinfo {pages} {9–15}\BibitemShut {NoStop}%
\bibitem [{Cod()}]{CodeRefactoring}%
  \BibitemOpen
  \href@noop {} {\enquote {\bibinfo {title} {Code refactoring},}\ }\bibinfo
  {howpublished} {\url{https://en.wikipedia.org/wiki/Code_refactoring}},\
  \bibinfo {note} {accessed November 28, 2023}\BibitemShut {NoStop}%
\bibitem [{\citenamefont {Wolfe}()}]{UniMem}%
  \BibitemOpen
  \bibfield  {author} {\bibinfo {author} {\bibfnamefont {M.}~\bibnamefont
  {Wolfe}},\ }\href@noop {} {\enquote {\bibinfo {title} {Openacc and cuda
  unified memory},}\ }\bibinfo {howpublished}
  {\url{https://www.pgroup.com/blogs/posts/openacc-unified-memory.htm}},\
  \bibinfo {note} {accessed June 8, 2023}\BibitemShut {NoStop}%
\bibitem [{\citenamefont {Valeev}(2016)}]{Valeev16}%
  \BibitemOpen
  \bibfield  {author} {\bibinfo {author} {\bibfnamefont {E.~F.}\ \bibnamefont
  {Valeev}},\ }\href@noop {} {\enquote {\bibinfo {title} {Libint: A library for
  the evaluation of molecular integrals of many-body operators over gaussian
  functions},}\ }\bibinfo {howpublished} {http://libint.valeyev.net/} (\bibinfo
  {year} {2016}),\ \bibinfo {note} {version 2.2.0}\BibitemShut {NoStop}%
\bibitem [{Reg()}]{RegressionTesting}%
  \BibitemOpen
  \href@noop {} {\enquote {\bibinfo {title} {Regression testing},}\ }\bibinfo
  {howpublished} {\url{https://en.wikipedia.org/wiki/Regression_testing}},\
  \bibinfo {note} {accessed November 28, 2023}\BibitemShut {NoStop}%
\bibitem [{\citenamefont {Williams}\ \emph {et~al.}(2009)\citenamefont
  {Williams}, \citenamefont {Waterman},\ and\ \citenamefont
  {Patterson}}]{Williams09}%
  \BibitemOpen
  \bibfield  {author} {\bibinfo {author} {\bibfnamefont {S.}~\bibnamefont
  {Williams}}, \bibinfo {author} {\bibfnamefont {A.}~\bibnamefont {Waterman}},
  \ and\ \bibinfo {author} {\bibfnamefont {D.}~\bibnamefont {Patterson}},\
  }\href {\doibase 10.1145/1498765.1498785} {\bibfield  {journal} {\bibinfo
  {journal} {Commun. ACM}\ }\textbf {\bibinfo {volume} {52}},\ \bibinfo {pages}
  {65–76} (\bibinfo {year} {2009})}\BibitemShut {NoStop}%
\bibitem [{\citenamefont {Yliluoma}()}]{GuideOpenMP}%
  \BibitemOpen
  \bibfield  {author} {\bibinfo {author} {\bibfnamefont {J.}~\bibnamefont
  {Yliluoma}},\ }\href@noop {} {\enquote {\bibinfo {title} {Guide into
  openmp},}\ }\bibinfo {howpublished}
  {\url{https://bisqwit.iki.fi/story/howto/openmp/}},\ \bibinfo {note}
  {accessed June 8, 2023}\BibitemShut {NoStop}%
\end{thebibliography}%

\end{document}